\documentclass{aa}  

\usepackage[colorlinks,linkcolor=blue,citecolor=blue]{hyperref}
\usepackage{graphicx} 
\usepackage{lscape}
\usepackage{txfonts}
\usepackage{booktabs}
\usepackage{ulem} 

\usepackage{natbib,twoopt}

\makeatletter
\newcommandtwoopt{\citeads}[3][][]{\href{http://adsabs.harvard.edu/abs/#3}%
    {\def\hyper@linkstart##1##2{}%
     \let\hyper@linkend\@empty\citealp[#1][#2]{#3}}}
  \newcommandtwoopt{\citepads}[3][][]{\href{http://adsabs.harvard.edu/abs/#3}%
    {\def\hyper@linkstart##1##2{}%
     \let\hyper@linkend\@empty\citep[#1][#2]{#3}}}
  \newcommandtwoopt{\citetads}[3][][]{\href{http://adsabs.harvard.edu/abs/#3}%
    {\def\hyper@linkstart##1##2{}%
     \let\hyper@linkend\@empty\citet[#1][#2]{#3}}}
  \newcommandtwoopt{\citeyearads}[3][][]%
    {\href{http://adsabs.harvard.edu/abs/#3}
    {\def\hyper@linkstart##1##2{}%
     \let\hyper@linkend\@empty\citeyear[#1][#2]{#3}}}
\makeatother

\begin{document}

\title{The eROSITA Final Equatorial-Depth Survey (eFEDS):}
\subtitle{X-ray properties and scaling relations of galaxy clusters and groups\thanks{The lists of X-ray observable measurements (Table~\ref{tab:R500}) and best-fit electron density model parameters (Table~\ref{tab:sb_fit}) of eFEDS clusters are available in electronic form at \url{https://erosita.mpe.mpg.de/edr/eROSITAObservations/Catalogues} and at the CDS via anonymous ftp to \url{dsarc.u-strasbg.fr} (\url{130.79.128.5}) or via \url{http://cdsarc.u-strasbg.fr/viz-bin/cat/J/A+A/661/A7}}}
\author{Y.~Emre~Bahar\inst{1}\thanks{e-mail: \href{mailto:ebahar@mpe.mpg.de}{\tt ebahar@mpe.mpg.de}},
Esra~Bulbul\inst{1},
Nicolas~Clerc\inst{2},
Vittorio~Ghirardini\inst{1},
Ang~Liu\inst{1},
Kirpal~Nandra\inst{1}, 
Florian~Pacaud\inst{3},
I-Non~Chiu\inst{4},
Johan~Comparat \inst{1},
Jacob~Ider-Chitham \inst{1},
Mathias~Klein \inst{5},
Teng~Liu \inst{1},
Andrea~Merloni \inst{1},
Konstantinos~Migkas\inst{3},
Nobuhiro~Okabe\inst{6,7,8},
Miriam~E.~Ramos-Ceja\inst{1},
Thomas~H.~Reiprich\inst{3},
Jeremy~S.~Sanders\inst{1},
Tim~Schrabback\inst{3}
}

\institute{
\inst{1}{Max Planck Institute for Extraterrestrial Physics, Giessenbachstrasse 1, 85748 Garching, Germany}\\
\inst{2}{IRAP, Université de Toulouse, CNRS, UPS, CNES, Toulouse, France}\\
\inst{3}{Argelander-Institut f{\"{u}}r Astronomie (AIfA), Universit{\"{a}}t Bonn, Auf dem H{\"{u}}gel 71, 53121 Bonn, Germany}\\
\inst{4}{Tsung-Dao Lee Institute, and Key Laboratory for Particle Physics, Astrophysics and Cosmology, Ministry of Education, Shanghai Jiao Tong University, Shanghai 200240, China}\\
\inst{5}Universitaets-Sternwarte Muenchen, Fakultaet fuer Physik, LMU Munich, Scheinerstr. 1, 81679 Munich, Germany \\
\inst{6}{Physics Program, Graduate School of Advanced Science and Engineering, Hiroshima University, 1-3-1 Kagamiyama, Higashi-Hiroshima, Hiroshima 739-8526, Japan}\\
\inst{7}{Hiroshima Astrophysical Science Center, Hiroshima University, 1-3-1 Kagamiyama, Higashi-Hiroshima, Hiroshima 739-8526, Japan}\\
\inst{8}{Core Research for Energetic Universe, Hiroshima University, 1-3-1, Kagamiyama, Higashi-Hiroshima, Hiroshima 739-8526, Japan}\\
}

\titlerunning{X-ray properties and scaling relations of the eFEDS galaxy clusters and groups}
\authorrunning{Bahar et al.}

\abstract
 {Scaling relations link the physical properties of clusters at cosmic scales. They are used to probe the evolution of large-scale structure, estimate observables of clusters, and constrain cosmological parameters through cluster counts.}
 {We investigate the scaling relations between X-ray observables of the clusters detected in the eFEDS field using Spectrum-Roentgen-Gamma/eROSITA observations taking into account the selection effects and the distributions of observables with cosmic time.}
 {We extract X-ray observables ($L_{\rm X}$, $L_{\rm bol}$, $T$, $M_{\rm gas}$, $Y_{\rm X}$) within $R_{500}$ for the sample of 542 clusters in the eFEDS field. By applying detection and extent likelihood cuts, we construct a subsample of 265 clusters with a contamination level of $<10\%$ (including AGNs and spurious fluctuations) to be used in our scaling relations analysis. The selection function based on the state-of-the-art simulations of the eROSITA sky is fully accounted for in our work.}
 {We provide the X-ray observables in the core-included $<R_{500}$ and core-excised $0.15R_{500}-R_{500}$ apertures for 542 galaxy clusters and groups detected in the eFEDS field. Additionally, we present our best-fit results for the normalization, slope, redshift evolution, and intrinsic scatter parameters of the X-ray scaling relations between $L_{\rm X}-T$, $L_{\rm X}-M_{\rm gas}$, $L_{\rm X}-Y_{\rm X}$, $L_{\rm bol}-T$, $L_{\rm bol}-M_{\rm gas}$, $L_{\rm bol}-Y_{\rm X}$, and $M_{\rm gas}-T$. We find that the best-fit slopes significantly deviate from the self-similar model at a $>4\sigma$ confidence level, but our results are nevertheless in good agreement with the simulations including non-gravitational physics, and the recent results that take into account selection effects.}
 {The strong deviations we find from the self-similar scenario indicate that the non-gravitational effects play an important role in shaping the observed physical state of clusters. This work extends the scaling relations to the low-mass, low-luminosity galaxy cluster and group regime using eFEDS observations, demonstrating the ability of eROSITA to measure emission from the intracluster medium out to $R_{500}$ with survey-depth exposures and constrain the scaling relations in a wide mass--luminosity--redshift range.
 }
 
\maketitle
\section{Introduction}
Galaxy clusters, which are formed by the gravitational collapse of the largest density peaks in the primordial density field, represent the largest virialized objects in the Universe. Embedded in the cosmic web, they evolve and grow through mergers and by accreting smaller subhaloes via the surrounding large-scale structure \citep[e.g.,][]{Kravtsov+12}. The number counts of clusters of galaxies as a function of redshift and their mass is a powerful cosmological probe that is orthogonal and complementary to other cosmological geometrical experiments (e.g., \citealt{Pillepich2012,Mantz2015,Schellenberger2017,Pacaud2018}, also see \citealt{Pratt2019} for a review). Additionally, based on the current Lambda cold dark matter ($\Lambda$CDM) cosmological model, galaxy clusters are among the structures formed last, and therefore capture the formation history and the growth of the structure in the Universe.
 
Well-established scaling relations between cluster mass and observables provide a way forward for cosmological investigations using clusters of galaxies. Accurate estimates of cluster total masses are crucial ingredients for exploiting the cluster number counts as cosmological probes. However, measurements of masses of individual clusters through multi-wavelength (X-ray, optical, weak lensing, and radio) observations can be expensive for larger cluster samples. Scaling relations aid this problem and bridge cluster number counts with cosmology.
On the other hand, the scaling relations between observables and their evolution allow us to constrain intracluster medium (ICM) physics and theoretical models based on gravitational collapse \citep[e.g.,][]{kaiser1986,Ascasibar2006,Short2010,Capelo2012}. \citet{kaiser1986} modeled the formation of clusters as scale-free collapses of initial density peaks and derived relations between ICM properties that result in clusters at different redshifts and masses being scaled versions of each other. This is called the self-similar model in the literature. Other nongravitational physical processes, such as radiative cooling, galactic winds, turbulence, and AGN feedback, that affect the formation and evolution of these objects throughout cosmic time may have imprints on these relations. In observational studies, these imprints are quantified by measuring deviations from the self-similar scaling relations. Clusters of galaxies owing to their deep potential well are less prone to these nongravitational processes, while the intra-group gas in galaxy groups can be significantly impacted by nongravitational physics \citep[e.g.,][]{Tozzi2001, Borgani2002, Babul2002, Puchwein2008, Biffi+14, Barnes2017}. 

The majority of the baryonic content of the clusters is in the form of X-ray-emitting hot ionized plasma, the ICM. Being in the fully ionized state and reaching up to $10^{8}$~Kelvin in temperature, the ICM emits primarily in X-rays through thermal Bremsstrahlung, offering an opportunity to measure physical properties of the ICM, to establish scaling relations between these properties and mass, and to constrain their evolution over cosmic time. The scaling relations between X-ray observables and mass have been extensively explored for massive clusters in the literature, selected in various ways by the  large-area, multi-wavelength surveys \citep[e.g.,][]{Mantz2010a, Bulbul2019}. However, samples including a sufficient number of uniformly selected groups covering the low-mass, low-redshift, and low-luminosity range with adequate count rates are limited. Studies of the scaling relations of galaxy groups and clusters spanning a wide mass, luminosity, and redshift range with large-area surveys with a well-understood selection will improve our understanding of the interplay between galaxy evolution, AGN feedback, and gravitational processes in these deep potential wells. XMM{\it-Newton}'s largest observational programme XXL \citep{Pierre2011} served as a bridge between narrow and deep observations \citep[e.g., CDF-S,][]{Finoguenov2015} and very wide, moderately deep observations \citep[e.g., RASS,][]{Ebeling1998} by populating the intermediate parameter space with detected clusters. Most recently, the extended ROentgen Survey with an Imaging Telescope Array \citep[eROSITA,][]{Merloni2012, Predehl2021} carried out its eROSITA Final Equatorial-Depth Survey (eFEDS) observations and provided numerous cluster detections that span a large mass--redshift space. eROSITA on board the Spectrum-Roentgen-Gamma (SRG) mission continues to detect large numbers of clusters spanning a wide range of redshift and mass since its launch in 2019. It will provide sufficient statistical power and place the tightest constraints on these scaling relations for probing their mass and redshift evolution.

\begin{figure*}
\begin{tabular}{c}
\includegraphics[width=0.995\textwidth,trim={0 0.4cm 0 0}, clip]{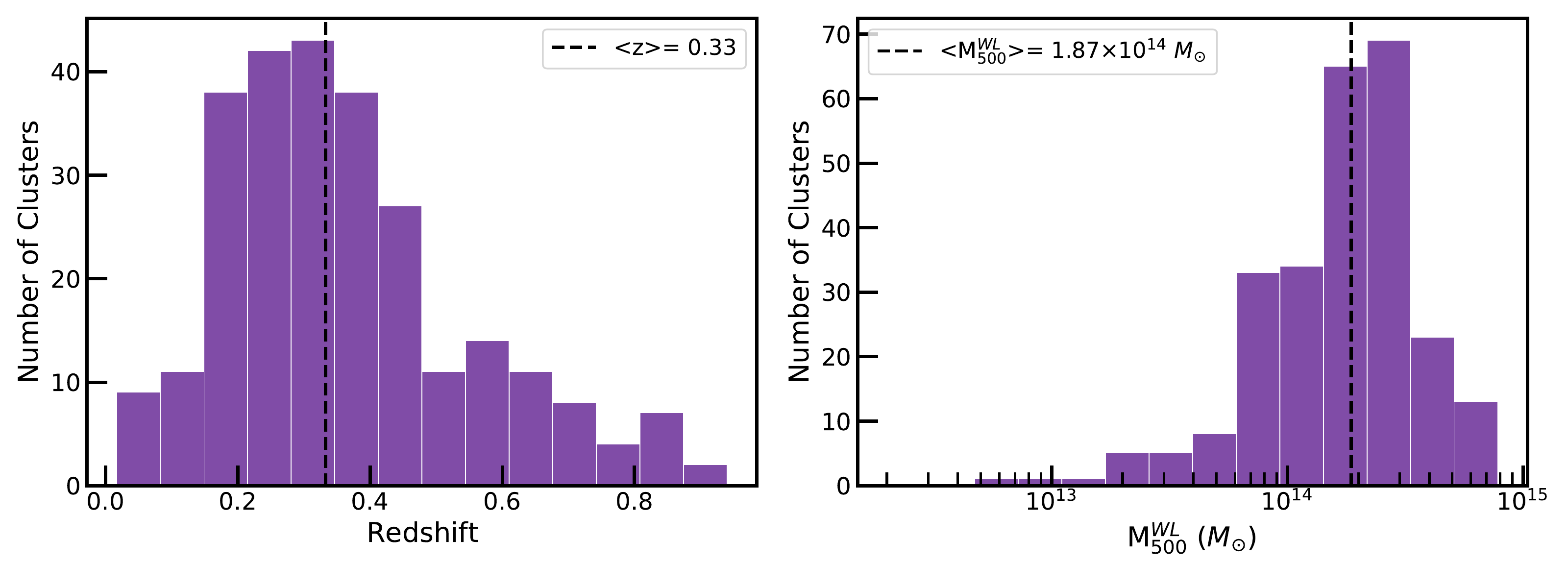} 
\end{tabular}
\caption{{\it Left}: Redshift histogram of the $\mathcal{L}_{\rm det} > 15$, $\mathcal{L}_{\rm ext} > 15$ sample used in this work. {\it Right}: Weak lensing calibrated total mass ($M_{500}^{\rm WL}$) histogram of the $\mathcal{L}_{\rm det} > 15$, $\mathcal{L}_{\rm ext} > 15$ sample used in this work excluding upper limit measurements. Medians of the measurements are marked with dashed line. The overdensity radii of $R_{500}$ within which we measure the X-ray observables are calculated from these mass measurements \citep[see][]{Chiu2021}.
\label{fig:hist}}
\end{figure*}

The eFEDS was performed during eROSITA's calibration and performance verification phase \citep{Predehl2021,Brunner2021,liu2021a}. eFEDS, the first (mini)survey of eROSITA, is designed to serve as a demonstration of the observational capabilities of the telescope to the scientific community. The survey area is located at (approximately) 126$^{\circ}$ < R.A. < 146$^{\circ}$ and $-3^{\circ}$ < Dec. < $+6^{\circ}$ and covers a total of $\sim$140 deg$^{2}$. The exposure time of the survey area is mostly uniform with average vignetted and unvignetted exposure times of $\sim1.3$ and $\sim2.2$~ks, respectively \citep{Brunner2021}. The eFEDS area is also covered in survey programs of other telescopes such as the Hyper Supreme-Cam Subaru Strategic Program \citep[HSC-SSP;][]{HSC-SSP}, DECaLS \citep[Dark Energy Camera Legacy Survey,][]{2019Dey}, SDSS \citep[Sloan Digital Sky Survey,][Ider-Chitham et al., in prep.]{SDSS}, 2MRS \citep[2MASS Redshift Survey,][]{2MRS}, and GAMA \citep[Galaxy And Mass Assembly,][]{GAMA}. These observations are used to optically confirm the detected clusters and measure their redshifts \citep[][Ider-Chitham et al., in prep.]{Klein2021}. In addition to the optical confirmation and redshift determination, HSC-SSP observations are also used in measuring the weak lensing mass estimates of the detected clusters. The observables presented in this work are measured using $R_{500}$\footnote{$R_{500}$ is the overdensity radius within which the density of the cluster is 500 times the critical density of the Universe at the cluster's redshift.} values inferred from these weak lensing measurements \citep{Chiu2021}. In this work, we provide X-ray properties of the 542 galaxy clusters and groups in the full eFEDS-extent-selected sample in two apertures ($r<R_{500}$ and $0.15R_{500}<r<R_{500}$) \citep{liu2021a}. Additionally, we investigate the scaling relations between core-included ($r<R_{500}$) X-ray observables in a subsample of 265 galaxy clusters and groups with a lower level of contamination by noncluster detections. This work expands the scaling relation studies to the poorly explored mass ($6.86 \times10^{12}~$M$_{\odot}<M_{500}< 7.79\times10^{14}$~M$_{\odot}$), luminosity ($8.64 \times 10^{40}~{\rm erg}~{\rm s}^{-1}<L_{\rm X}<3.96 \times 10^{44}~{\rm erg}~{\rm s}^{-1}$), and redshift ($0.017<z<0.94$) ranges with the largest number of galaxy groups and clusters, paving the way for similar studies using the eROSITA All-Sky survey (eRASS) observations. We note that the scaling relations between X-ray observables and weak lensing masses have already been published in our companion paper, \citet{Chiu2021}. The selection function is based on the realistic full sky simulations of eROSITA and is fully accounted for in our results \citep{2020Comparat}. Throughout this paper, the best-fitting thermal plasma temperature to the cluster spectra is marked with $T$, $L_{\rm X}$ stands for soft-band X-ray luminosity calculated in the $0.5-2.0$~keV energy band, $L_{\rm bol}$ stands for the bolometric luminosity calculated in the $0.01-100$~keV energy band, the errors correspond to $68\%,$ and  we adopt a flat $\Lambda$CDM cosmology using the \citet{Planck2016} results, namely $\Omega_m =0.3089$, $\sigma_8=0.8147,$ and $H_0 = 67.74$~km~s$^{-1}$~Mpc$^{-1}$

\section{Data analysis}
\label{sec:analysis}
\subsection{Data reduction and sample selection}

The eFEDS observations were performed by eROSITA between  4 and 7 November 2019. The observation strategy allowed the eFEDS field to be surveyed nearly uniformly with a vignetted exposure of $\sim1.3$~ks which is similar to the expected vignetted exposure of the final eROSITA All-Sky Survey (eRASS8) at the equatorial regions. Initial processing of the eFEDS observations was carried out using the eROSITA Standard Analysis Software System \citep[{\tt eSASS}, version {\tt eSASSusers\_201009},][]{Brunner2021}. 
In this paper, we only present an outline of the summary of the data reduction and source detection. We refer the reader to \citet{Brunner2021} and \citet{liu2021a} for a more detailed explanations of these steps. We first applied filtering to X-ray data, removing dead time intervals and frames, corrupted events, and bad pixels. 
Images created in the $0.2-2.3$~keV band using all available telescope modules (TMs) are passed to eSASS source-detection tools in order to perform the source detection procedure and provide extension and detection likelihoods. After applying a detection likelihood ($\mathcal{L}_{\rm det}$) threshold of 5 and an extension likelihood ($\mathcal{L}_{\rm ext}$) threshold of 6, we obtained 542 cluster candidates in the eFEDS field \citep{Brunner2021}. The physical properties of these clusters, such as soft-band and bolometric luminosities, and ICM temperature measurements within a physical radius of 300~kpc and 500~kpc are provided by \citet{liu2021a}.

We used realistic simulations of the eFEDS field \citep{LiuT2021} in order to measure the contamination fractions of samples with different $\mathcal{L}_{\rm det}$ and $\mathcal{L}_{\rm ext}$ cuts. According to these simulations, the eFEDS cluster catalog, which consists of 542 clusters, has a contamination fraction of $\sim$20\%. This is a relatively high contamination rate for statistical studies. In order to avoid  significant bias caused by the noncluster sources present in the sample (e.g., AGNs and spurious sources), we applied $\mathcal{L}_{\rm det} > 15$ and $\mathcal{L}_{\rm ext} > 15$ cuts that give us a sample of 265 clusters with an expected contamination fraction of 9.8\%. The final sample covers  a total mass range of ($6.86 \times10^{12}~$M$_{\odot}<M_{500}< 7.79\times10^{14}$~M$_{\odot}$, a luminosity range of $8.64 \times 10^{40}~{\rm erg}~{\rm s}^{-1}<L_{\rm X}<3.96 \times 10^{44}~{\rm erg}~{\rm s}^{-1}$, and a redshift range of ($0.017<z<0.94$). The redshift and mass histograms of this final subsample are shown in Fig~\ref{fig:hist}. Consisting of 68 low-mass ($<10^{14}$~M$_{\odot}$) galaxy groups, this work extends the scaling relation studies to the low-mass range with one of the largest group samples detected uniformly to date.
\subsection{X-ray observables within $R_{500}$}

One of the main goals of this paper is to provide X-ray properties of eFEDS clusters within the overdensity radius of $R_{500}$. Here we provide a short summary of the methods we employed to extract X-ray observables. For a complete description, we refer the reader to \citet{Ghirardini2021} and \citet{liu2021a}.

The measurements of $R_{500}$ used in this work are obtained from the weak lensing calibrated cluster masses presented in our companion paper \citep{Chiu2021}. Calibration is applied by using the eFEDS observations of the same cluster sample used in this work which enables the $R_{500}$ measurements to be self-consistent. Mass estimates are obtained by jointly modeling the eROSITA X-ray count-rate ($\eta$) and HSC shear profile ($g_{+}$) as a function of cluster mass ($M_{500}$) and obtaining a scaling relation between $\eta-M_{500}-z$. After obtaining the mass estimates, $R_{500}$ measurements are calculated by $R_{500}=\left( \frac{3  }{  4\pi} \frac{M_{500}}{500\rho_{c}}\right )^{1/3}$ where $\rho_c$ is the critical density at a given redshift and cosmology. We refer the reader to \citet{Chiu2021} for a more detailed description of the HSC weak-lensing mass calibration analysis.

X-ray spectra of clusters are extracted within $R_{500}$, both core-included ($r<R_{500}$) and core-excised ($0.15R_{500}<r<R_{500}$), using the \texttt{eSASS} code {\tt srctool}. The background spectra are extracted from an annular region that is $4-6~R_{500}$ away from the clusters' centroid. 
We fit the X-ray spectra with an absorbed {\tt apec} thermal plasma emission model \citep{smith2001,2012Foster} to represent the ICM emission. The fitting band of $0.5-8$~keV was used for TMs 1, 2, 3, 4, and 6 and a more restricted band of $0.8-8$~keV was used for TMs 5 and 7 in the spectral fits due to the light leak noticed during the commissioning phase \citep[see][]{Predehl2021}. The Galactic hydrogen absorption is modeled using {\tt tbabs} \citep{2000Wilms}, where the column density $n_{\rm H}$ used is fixed to $n_{\rm H, tot}$ \citep{2013Willingale}, estimated at the position of the cluster center. The metallicity of the clusters is fixed to 0.3~$Z_{\odot}$, adopting the solar abundance table of \citet{asplund2009}. The background spectra and spectra are simultaneously fit to account for the background in the total spectra as described in detail by \citet{Ghirardini2021}. The background spectra are modeled with a set of {\tt apec} and power-law models representing instrumental background based on the filter-wheel closed data \citep[see ][]{Freyberg+20}\footnote{https://erosita.mpe.mpg.de/edr/eROSITAObservations/EDRFWC/}, cosmic background (local bubble, galactic halo, and emission from unresolved AGNs). The best-fit values and standard deviations of the ICM temperatures ($T$) are measured using the Markov chain Monte Carlo (MCMC) method within XSPEC (version 12.11.0k).

We extract images and exposure maps in the $0.5-2.0$~keV energy band to obtain cluster density profiles. We model the two-dimensional distribution of photons by projecting the \citet{Vikhlinin+06} density model. Point sources are either modeled or masked depending on their fluxes; see \citet{Ghirardini2021b} for further details. The cosmic background contribution is added to the total model as a constant. The resulting total image is finally convolved with eROSITA's vignetted exposure map, while the instrumental background model is folded with the unvignetted exposure map. A Poisson log-likelihood in MCMC is used to estimate best-fit cluster model parameters. Finally, the electron density ($n_e$) profile of the gas is obtained by measuring the emissivity using the temperature information recovered from the spectral analysis. Best-fit parameters of clusters to the \citet{Vikhlinin+06} electron density profile model are presented in Table~\ref{tab:sb_fit}. In order to obtain luminosity profiles, $L_{\rm X}(r)$ and $L_{\rm bol}(r)$, we calculated conversion factors from count rate to luminosity in soft ($0.5-2.0$~keV) and bolometric ($0.01-100$~keV) energy bands.

The gas mass (or ICM mass) of the clusters enclosed within $R_{500}$ is computed by integrating the gas electron density assuming spherical symmetry:

\begin{equation}
M_{\rm g} = \mu_e m_p \int_0^{R_{500}} n_e (r)  4 \pi r^2 dr  
\label{eq:Mg}
,\end{equation}
\noindent where $n_e$ is the number density of electrons, m$_p$ is the proton mass, and $\mu_e = 1.1548$ is the mean molecular weight per electron calculated using the \citet{asplund2009} abundance table \citep{Bulbul2010}.
Lastly, $Y_{\rm X}$  is calculated by multiplying the gas mass ($M_{\rm gas}$) with the gas temperature ($T$) as
\begin{equation}
Y_{\rm X} = M_{\rm gas} \cdot T
\label{eq:Yx}
,\end{equation}
which is introduced by \citet{Kravtsov2006} as a low scatter mass estimator.

We note that in our analysis, uncertainties in $R_{500}$ measurements are fully propagated using the MCMC chains and the redshift errors are neglected. We use the single temperature in our calculations as the survey data do not have sufficient depth to recover the temperature profiles as a function of radius. For all eFEDS clusters, we provide the core-included ($r<R_{500}$) X-ray observables within $R_{500}$, including $T$, $L_{\rm X}$, $L_{\rm bol}$, $M_{\rm gas}$, and $Y_{\rm X}$ as well as the core-excluded X-ray observables extracted between $0.15R_{500}-R_{500}$ ($T_{\rm cex}$, $L_{\rm X,cex}$, $L_{\rm bol,cex}$) in Table~\ref{tab:R500}. eROSITA's field-of-view-averaged point spread function (PSF) half-energy width is $\sim26\arcsec$ which is comparable to the cores ($0.15R_{500}$) of the majority of clusters. This has a mild effect on the $L_{\rm X,cex}$ measurements because we deconvolve the surface brightness profiles with the PSF and use the best-fit core-included temperatures for the emissivity. However, given the limited photon statistics, only a first-order PSF correction is applied to the $T_{\rm cex}$ measurements where the flux changes at different energies are compensated by assuming the spectrum to be similar over the whole of the source. Therefore, we advise the reader to approach $T_{\rm cex}$ measurements with caution.

\begin{figure}
\centering
\begin{tabular}{c}
\includegraphics[width=0.49\textwidth]{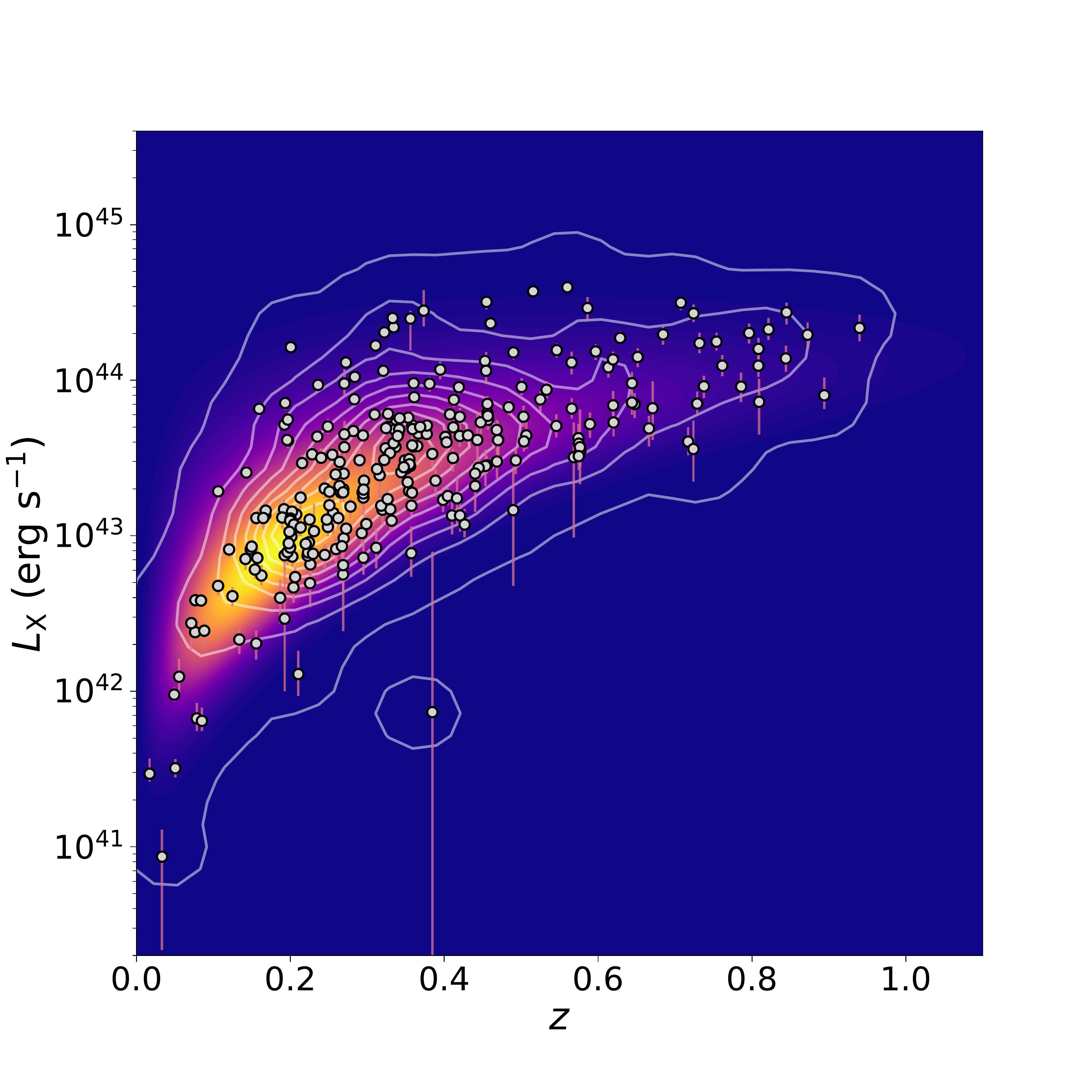} 
\end{tabular}
\caption{Soft band ($0.5-2.0$ keV) X-ray luminosity and redshift measurements of the clusters in the eFEDS field that satisfies the $\mathcal{L}_{\rm det} > 15$ and $\mathcal{L}_{\rm ext} > 15$ condition. Luminosities are measured within apertures of $R_{500}$. White solid curves are smoothed contours of the $L_{\rm X}-z$ data points and the color is proportional to the PDF of the hypothetical $L_{\rm X}-z$ distribution modeled as $P(I,L_{\rm X},z)= P(I|L_{\rm X},z) P(L_{\rm X}|z) P(z)$ (see Sect.~\ref{sec:selectionfunction} for the description of the model).
\label{Ldists}}
\end{figure}

In this work, we focus on the scaling relations between X-ray observables, namely $L-T,\ L-M_{\rm gas},\ L-Y_{\rm X}$, and  $\ M_{\rm gas}-T$. The scaling relations between observables and cluster mass ($M_{500}$) obtained from weak-lensing observations are already provided in the companion paper by \citet{Chiu2021}.
Although we provide measurements of the core-excluded observables in this paper in Table~\ref{tab:R500}, we only use the core-included observables in our further analysis for the scaling relations. The reasons for this are twofold, and are related to the selection function, and the decrease in the statistics. Our selection function is built using the core-included observables from the simulations of eROSITA sky \citep{2020Comparat}. Constructing selection functions with the core-excised observables relies on modeling the PSF accurately in simulations. Our imaging analysis and spectral fits account for the PSF spilling,but this analysis is not available yet in simulations. As a workaround, one could model the relation between the core-excised and core-included observables (e.g., $P(Y_{\rm X,cex} | Y_{\rm X})$), but a significant fraction of eFEDS clusters populate a previously poorly explored parameter space and such an approach requires a good understanding of the surface brightness profiles of these clusters. Secondly, when the core is excised, the temperature measurements become either loose or lost due to the decrease in photon statistics. This affects the reliability of the X-ray observable measurements used in our fits and may lead to biased constraints on the scaling relations. A full analysis with the core-excised observables will be carried out for the clusters detected in the eRASS observations, where we expect to have a larger sample of clusters with a higher depth around the ecliptic poles (Ghirardini et al. 2022, in prep.).

\section{Modeling and fitting of the scaling relations}
\label{sec:modelingandfitting}
We model the scaling relations and the likelihoods for different pairs of observables in a similar manner with minor tweaks. Therefore, in this section, we present the general form of the scaling relations and the structure of the likelihood for two hypothetical observables: $X$ and $Y$.

\subsection{General form of the scaling relations}
\citet{kaiser1986} derived simple forms of scaling relations, namely self-similar relations, by assuming gravitational interactions to be the driving force of the evolution of groups and clusters. These relations suggest that the observables of clusters follow these simple power-law relations. Departures from these relations are often interpreted as a result of non-gravitational physical processes, such as radiative cooling, galactic winds, and AGN feedback that can have a significant impact on the distribution of baryons in the 
ICM and energy budget of the system \citep[][]{Bhattacharya2008, McCarthy2010, Fabjan2010, Bulbul2016, Giodini2013, Lovisari2020}.

\begin{table}
\centering
\caption{Self-similar expected model parameter values of scaling relations of the form $Y = A~Y_{\rm piv}  \left (\frac{\rm X}{X_{\rm piv}} \right )^B \left (\frac{E(z)}{E(z_{\rm piv})} \right )^C$}
\label{tab:selfsimilar}
\begin{tabular}{cccc}
\toprule\midrule
\multicolumn{2}{c}{Relation} &  &  \\
\cmidrule(lr){1-2}
$Y$ & $X$ & $B_{\rm self}$ & $C_{\rm self}$ \\
\midrule
$L_{\rm X}$ & $T$ & 3/2 & 1 \\
$L_{\rm X}$ & $M_{\rm gas}$ & 1 & 2 \\
$L_{\rm X}$ & $Y_{\rm X}$ & 3/5 & 8/5 \\
\midrule
$L_{\rm bol}$ & $T$ & 2 & 1\\
$L_{\rm bol}$ & $M_{\rm gas}$ & 4/3 & 7/3 \\
$L_{\rm bol}$ & $Y_{\rm X}$ & 4/5 & 9/5 \\
\midrule
$M_{\rm gas}$ & $T$ & 3/2 & -1 \\
\bottomrule
\end{tabular}
\end{table}

In this work, we use a relation that takes into account the power-law dependence and the redshift evolution of the form

\begin{equation}\label{eq0}
Y = A~Y_{\rm piv}  \left (\frac{\rm X}{X_{\rm piv}} \right )^B \left (\frac{E(z)}{E(z_{\rm piv})} \right )^C
,\end{equation}
\noindent where $Y_{\rm piv}$, $X_{\rm piv}$, and $z_{\rm piv}$ are the pivot values of the sample, and $A$, $B$, and $C$ are the normalization, power-law slope, and  redshift evolution exponent, respectively. The redshift evolution is modeled using the evolution function which is defined as $E(z)=H(z)/H_0$ where $H(z)$ is the Hubble-Lema\^{i}tre parameter and $H_{0}$ is the Hubble constant.

\subsection{Likelihood}
\label{sec:likelihood}
In our fits to the scaling relations, we take into account various observational and physical effects by adding the relevant components to the corresponding likelihood function similar to the method presented in \citet{Giles2016} for the XXL clusters. The joint probability function in terms of the measured values ($\hat{X}$, $\hat{Y}$) of the true values of the observables $X$ and $Y$ is given by

\begin{equation}\label{eq1}
P(\hat{Y},\hat{X}, Y, X,I| \theta,z) = P(I|Y, z)P(\hat{Y},\hat{X} | Y, X) P (Y | X,\theta, z) P(X | z), 
\end{equation}

\noindent where $P(I|Y, z)$, also known as the selection function, is the probability of a cluster being included ($I$) in our sample, $P(\hat{Y},\hat{X} | Y, X)$ is the two-dimensional measurement uncertainty, $P (Y | X,\theta, z)$ is the modeled $Y-X$ relation, and the $P(X | z)$ term is the cosmological distribution of the observable $X$. The variable $\theta$ in the scaling relation term marks the free parameters of the scaling relation, such as $A$, $B$, $C$, and the scatter $\sigma_{Y|X}$. We note that in this work, correlations between the measurement uncertainties of observables $X$ and $Y$ are fully considered using the MCMC chains. We also note that the cosmological parameters are frozen throughout our analysis. More than $65\%$ of the clusters in our sample have spectroscopic redshifts and the remaining clusters have photometric redshift measurements using the high signal-to-noise-ratio HSC data, which provides uncertainties of the order of $0.3\%$ \citep[see][Ider-Chitham et al., in prep.]{Klein2021}. Therefore, we assume that the errors on the redshifts have negligible effects on our measurements, that is, $z=\hat{z}$. The variance in exposure time due to the overlapping regions and the missed observations due to malfunctions of telescope modules (TMs) \citep[see][for details]{Brunner2021} are accounted for by using the exposure time ($t_{\rm exp}$) at the X-ray center of each cluster when calculating $P(I|Y, z)$.

We model the $Y-X$ relation such that the observable $Y$ is distributed around the power-law scaling relation log-normally. Assumption of the log-normal distribution of X-ray observables is widely used in the literature \citep[e.g.,][]{Pacaud2007,Giles2016, Bulbul2019, Bocquet2019}. The scaling relation term  $P (Y | X,\theta, z)$ in Eq.~\ref{eq1} then becomes

\begin{equation}
\label{eqn:lognormalmodel}
P (Y | X,\theta, z) = \mathcal{LN} \left (\mu = A~Y_{\rm piv} \left (\frac{\rm X}{X_{\rm piv}} \right )^B \left (\frac{E(z)}{E(z_{\rm piv})} \right )^C , \sigma = \sigma_{Y|X} \right ).
\end{equation}

To obtain the cosmological distribution of the observable $X$ ($P(X|z)$), that is, the expected distribution of $X$ as a function of redshift given a fixed cosmology and an assumed $X-M$ scaling relation, we convert the \citet{tinker08} mass function to a Tinker $X$ function using the \citet{Chiu2021} weak-lensing mass-calibrated scaling relations obtained from the same cluster sample consistently. This conversion is applied such that the intrinsic scatter of the $X-M$ relation is taken into account by the following equation:

\begin{equation}
\label{eqn:xfunction}
P (X |\theta_{\rm WL}, z) = \int_M P(X|M,\theta_{\rm WL},z) P(M|z) dM
,\end{equation}
\noindent where $\theta_{\rm WL}$ is the best-fit result of the weak-lensing mass-calibrated scaling relation $X-M_{500}$. We note that the form of the $X-M$ relation presented in \citet{Chiu2021} is different than the form we use in our $Y-X$ relation. Hereafter, we do not include the $\theta_{\rm WL}$ term in $P (X |\theta_{\rm WL}, z)$, because it is fixed throughout the analysis. After properly defining all the terms in the joint distribution in Eq.~\ref{eq1}, we marginalize over the nuisance variables ($X$, $Y$) in order to get the likelihood of obtaining the measured observables ($\hat{X}$, $\hat{Y}$, $I$). The final likelihood of a single cluster then becomes

\begin{equation}
\begin{split}
P(\hat{Y},\hat{X}, I| \theta,z) = &\int \int_{Y,X} P(I|Y,z) \\ & \times P(\hat{Y},\hat{X} | Y, X) P (Y | X,\theta, z) P(X | z) dY dX.
\end{split}
\label{eqn:final_likelihood}
\end{equation}

To avoid significant bias in the results due to the assumed cosmological model and the exact form of the $X-M$ relation, we do not use the observed number of detected clusters as data, but instead we take it as a model parameter. In the Bayesian framework, this corresponds to using a likelihood that quantifies the probability of measuring $\hat{X_i}$ and $\hat{Y_i}$ observables given that the cluster is detected. Such a likelihood can be obtained using the Bayes theorem where the likelihood for the $i$th cluster becomes

\begin{equation}\label{eqn_yeni}
\mathcal{L}(\hat{Y}_i,\hat{X}_i| I, \theta,z_i) = \frac{ P(\hat{Y}_i,\hat{X}_i, I| \theta,z_i)}{\int \int_{\hat{Y}_i,\hat{X}_i} P(\hat{Y}_i,\hat{X}_i, I| \theta,z_i) d\hat{Y}_i d\hat{X}_i}
.\end{equation}

Lastly, the overall likelihood of the sample is obtained by multiplying the likelihoods of all clusters

\begin{equation}\label{eqn4}
\mathcal{L}(\hat{Y}_{\rm all},\hat{X}_{\rm all}|I, \theta,z) = \prod_{i}^{\hat{N}_{\rm det}} \mathcal{L}(\hat{Y}_i,\hat{X}_i|I, \theta,z_i)
,\end{equation}
\noindent where $\hat{Y}_{\rm all}$ and $\hat{X}_{\rm all}$ are the measurement observables of all clusters in the sample and $\hat{N}_{\rm det}$ is the number of detected clusters in our sample. 

This form of the likelihood is similar to those used in the literature; see for example \citet{Mantz2010b}. The most fundamental difference is the goal of this work, which is to fit the scaling relations at a fixed cosmology rather than simultaneously fitting scaling relations and cosmological parameters. Using this likelihood allows us to avoid including terms that have a strong dependence on cosmology, such as those in
\citet{Mantz2010a}, namely the probability of not detecting the model-predicted, undetected clusters, $P(\Bar{I}|\theta)$, possible ways of selecting $\hat{N}_{\rm det}$ clusters from the total sample $N$, ${{N}\choose{\hat{N}_{\rm det}}}$, and the prior distribution of the total number of clusters in the field, $P(N)$ \citep[see][for the use of these parameters]{Mantz2019}. Another benefit of using this likelihood is it allows the results to be less sensitive to the accuracy of the normalization of the $X-M$ relation and therefore makes our analysis more robust for the goal of this work.

\begin{table}
\centering
\caption{Median values of observables measured for the $\mathcal{L}_{\rm det} > 15$, $\mathcal{L}_{\rm ext} > 15$ clusters.}
\label{tab:pivots}
\begin{tabular}{cl}
\toprule\midrule
Parameters   &  Median/Pivots  \\
\midrule
$L_{\rm X}$ & $3.20 \times 10^{43}~{\rm erg}~{\rm s}^{-1}$ \\
$L_{\rm bol}$  & $9.49 \times 10^{43}~{\rm erg}~{\rm s}^{-1}$ \\
$T$ & 3.26 keV\\
$M_{\rm gas}$ & $1.04 \times 10^{13}~$M$_{\odot}$\\
$Y_{\rm X}$ & $3.75 \times 10^{13}~$M$_{\odot}~{\rm keV}$ \\
$z$ & 0.33 \\
\bottomrule
\end{tabular}
\tablefoot{These values are used as pivot values of observables in Eq.~\ref{eq0}}
\end{table}
\subsection{Modeling the selection function}
\label{sec:selectionfunction}
The selection function model adopted here, $P(I|Y, z)$ in Eq.~\ref{eqn:final_likelihood}, is similar to that described in \citet{liu2021a}. It relies on multiple mock realizations of the eFEDS field \citep{LiuT2021}. The simulations faithfully reproduce the instrumental characteristics of eROSITA and features induced by the scanning strategy (exposure variations, point-spread function, effective area, and the grasps of the seven telescopes.) Realistic foreground and background source models are associated with a full-sky light-cone N-body simulation assuming the Planck-CMB cosmology. These sources include stars, active galactic nuclei (AGN), and galaxy clusters. The method to associate AGN spectral templates to sources is derived from abundance-matching techniques. For clusters and groups, the association between a massive dark matter halo and an emissivity profile drawn from a library of observed templates depends on the mass, redshift, and dynamical state of the halo. In particular, relaxed halos are associated with gas distributions with higher central projected emission measures. The steps leading to the AGN and cluster simulations are extensively described in \citet{2019Comparat, 2020Comparat}. The {\sc SIXTE} engine \citep{2019Dauser} serves in converting sources into event lists, while the eSASS software \citep{Brunner2021} is used to process those lists and to deliver source catalogs.

The next steps are identical to those in \citet{liu2021a}, except for the definition of an extended detection which assumes $\mathcal{L}_{\rm det} > 15$ and $\mathcal{L}_{\rm ext} > 15$. In particular, pairs of the simulated and detected sources are looked for in the plane of the sky, accounting for their relative positions, their extents, and favoring association between bright sources in cases of ambiguity. Securely identified matches are flagged as a successful detection. 

The modeling of the detection probabilities involves interpolation across the multi-dimensional parameter space describing galaxy cluster properties, which includes their intrinsic soft-band or bolometric luminosity, their redshift, the local exposure time, and optionally the central emission measure. Other parameters are marginalized over, making the assumption that their distributions are correctly reflected in the simulations.
To this end, we make use of Gaussian Process classifiers, a class of nonparametric models which capture the variations of the detection probability under the assumption that the covariance function (kernel) is a squared exponential function. One advantage of using such models rather than the multi-dimensional spline interpolation, for example, is a more appropriate mathematical treatment of uncertainties, particularly in poorly populated areas of the parameter space. Two-thirds of the simulated clusters are used for training the classifiers, and the remaining third provides the material to test the performance of the classifiers and to assess their behavior on a realistic population of halos. In particular, we check that systems assigned a given detection probability by the classifier display a detection rate with a value close to that probability; in such cases the classifier is said to be well-calibrated.

These models are designed to emulate the whole chain of computationally expensive steps needed in performing an eFEDS end-to-end simulation \citep{LiuT2021}. It is worth noting that such selection functions have a range of applicability that is set by the simulation. 

In order to demonstrate the representativeness of the selection function, we model the luminosity distribution of the $\mathcal{L}_{\rm det} > 15$, $\mathcal{L}_{\rm ext} > 15$ clusters and compare it with the observed cluster distribution. We model it as $P(I,L_{\rm X},z)= P(I|L_{\rm X},z) P(L_{\rm X}|z) P(z)$ where we calculate $P(L_{\rm X}|z)$ using Eq.~\ref{eqn:xfunction} and the best-fit $L_{\rm X}-M$ relation presented in \citet{Chiu2021}. For the redshift distribution, we assume the comoving cluster density to be constant within our redshift span ($0<z<0.9$) so that $P(z)$ is proportional to the comoving volume shell $dV_{\rm c}(z)=\frac{c}{H_{0}}\frac{(1+z)^{2}{d_{\rm A}(z)}^{2}}{E(z)} \Omega_{\rm s} dz$ \citep{Hogg1999} where $c$ is the light speed, $H_{0}$ is the Hubble constant, $d_{\rm A}(z)$ is the angular diameter distance, and $\Omega_{\rm s}$ is the solid angle of the eFEDS survey. A comparison between the distribution of the luminosity measurements for the cluster sample with $\mathcal{L}_{\rm det} > 15$, $\mathcal{L}_{\rm ext} > 15$ selection and our model predicted by our selection function is shown in Fig.~\ref{Ldists}. The figure visually demonstrates the consistency of the luminosity distribution with redshift predicted from the selection function (plotted as the background color), and measurements from the eFEDS data (white data points and white contours). 

\begin{figure*}
\centering
\begin{tabular}{c}
\includegraphics[width=\textwidth]{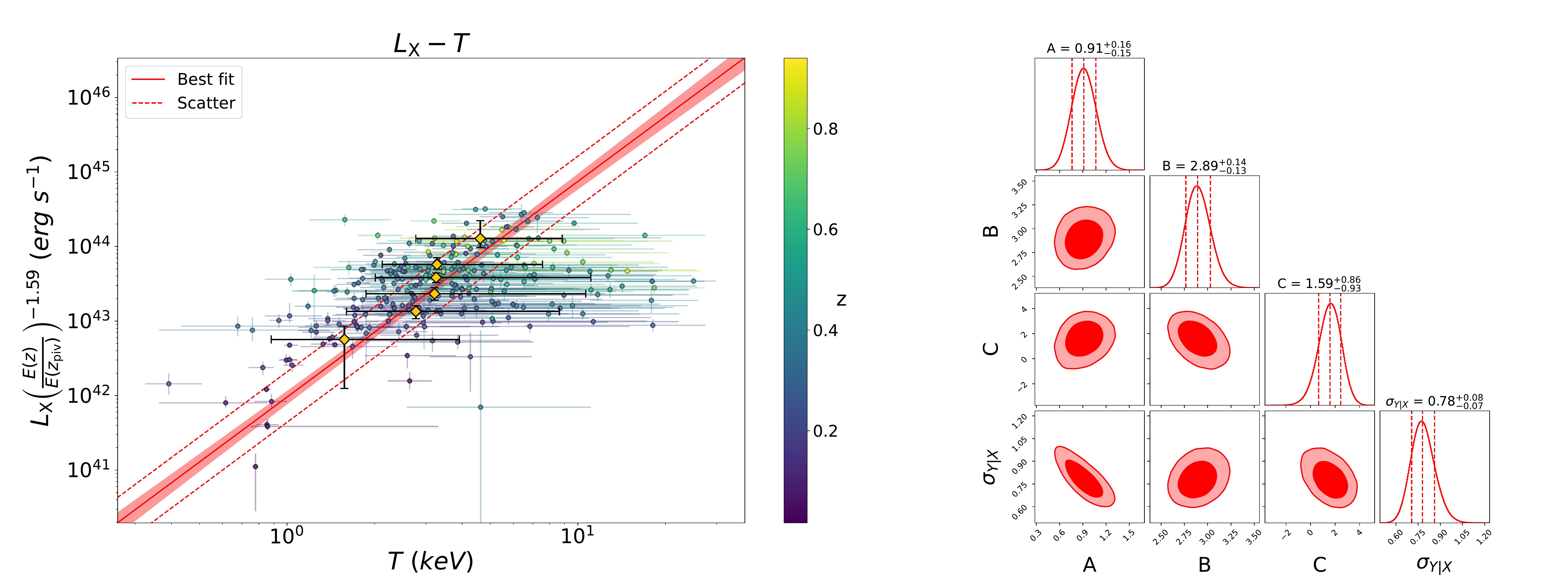} \\
\includegraphics[width=\textwidth]{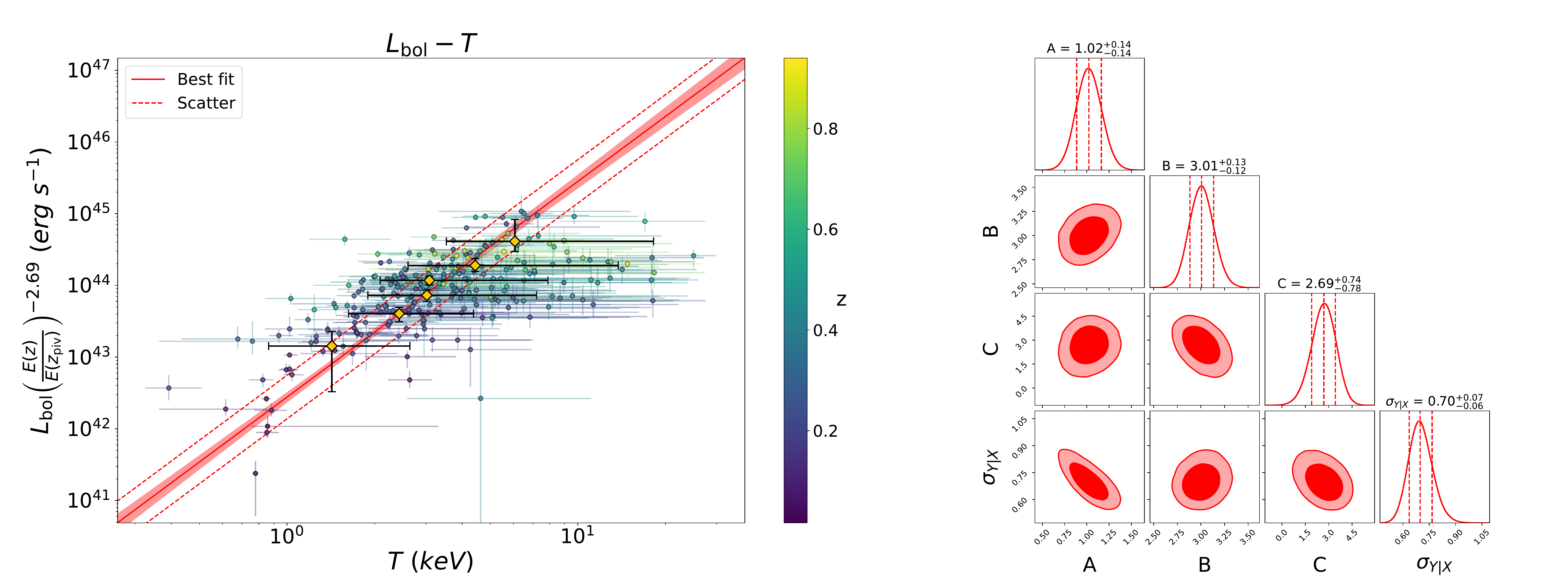}
\end{tabular}
\caption{$L-T$ scaling relations and the posterior distributions of the scaling relations parameters. {\it Left:} Soft band ($0.5-2.0$ keV) X-ray luminosity ($L_{\rm X}$), bolometric ($0.01-100$ keV) luminosity  ($L_{\rm bol}$), temperature ($T$), and redshift ($z$) measurements of the $\mathcal{L}_{\rm det} > 15$, $\mathcal{L}_{\rm ext} > 15$ sample and the best-fit scaling relation models. The light red shaded area indicates the 1$\sigma$ uncertainty of the mean of the log-normal model (see Eq.~\ref{eqn:lognormalmodel}) and the dashed red line indicates the best-fit standard deviation ($\sigma_{L|T}$) around the mean. Orange diamonds indicate median temperature measurements obtained from clusters between luminosity quantiles. {\it Right:} Parameter constraints of the $L_{\rm X}-T$ and $L_{\rm bol}-T$ relations obtained from the second half of the MCMC chains. Marginalized posterior distributions are shown on the diagonal plots and the joint posterior distributions are shown on off-diagonal plots. Red dashed vertical lines indicate the 32th, 50th, and 68th percentiles and contours indicate 68\% and 95\% credibility regions.
\label{figLT}}
\end{figure*}
\subsection{Fitting}
\label{sec:fitting}
We fit scaling relations using the MCMC sampler package {\tt emcee} \citep{emcee} with a likelihood described in Sect.~\ref{sec:likelihood}. Before we fit the real data, we validate our fitting code on simulated clusters. For the tests, we mock X-ray observables of a sample of 265 clusters corresponding to the same number of clusters in the sample selected with the criteria of $\mathcal{L}_{\rm det} > 15$ and $\mathcal{L}_{\rm ext} > 15$. Using the observed redshifts as priors, we sample the observables, $X$ and $Y$, from a bivariate distribution of the form
\begin{equation}\label{simeq}
P(Y, X, I| \theta_{\rm sim},z) = P(I|Y, z) P (Y | X,\theta_{\rm sim}, z) P(X | z), 
\end{equation}

\noindent where $P (Y | X,\theta_{\rm sim}, z)$ is the scaling relation term including intrinsic scatter and $\theta_{\rm sim}$ is the input scaling relation parameters for the simulated clusters. We then scatter the $X$ and $Y$ observables to mimic observational uncertainty and assign conservative error bars to model our observable measurements. We then run our fitting code on the simulated clusters with 100 walkers for 10000 steps and compare the best-fit $\theta$ values with the input parameters ($\theta_{\rm sim}$). We find that the fitting code successfully recovers all input parameters with a deviation within one sigma validating the performance and the accuracy of the code.

After the test run, we fit the X-ray scaling relations using the eFEDS measurements using  flat priors for all scaling relation parameters; $\mathcal{U}(-4,4)$ for the normalization ($A$), $\mathcal{U}(-10,10)$ for the slope ($B$), $\mathcal{U}(-10,10)$ for the redshift evolution exponent ($C$), and $\mathcal{U}(0.1,3.0)$ for the scatter ($\sigma_{\rm Y|X}$). The median values of the observables are used as the pivot values in our fits and are provided in Table~\ref{tab:pivots}.

In total, we perform two fits for each scaling relation. The first fits are performed with free redshift evolution exponents, $C$, and in the second fits the parameter $C$ is fixed to the self-similar values. The self-similar expectations are given in Table~\ref{tab:selfsimilar} for all scaling relations used in this work. The best-fit parameters of these seven relations can be found in Table~\ref{tab:scalingresults}. We provide our results and comparisons with the literature in Section~\ref{sec:results}. 

\begin{table*}
\centering
\caption{Best-fit parameters of the scaling relations}
\renewcommand{\arraystretch}{1.4}
\begin{tabular}{ c c c c c c c c c c c }
\toprule\midrule
 \multicolumn{2}{c}{}       & \multicolumn{8}{c}{Best fit parameters} \\
 \multicolumn{2}{c}{Relation}       & \multicolumn{4}{c}{Free redshift evolution} & \multicolumn{4}{c}{Self-similar redshift evolution} \\

\cmidrule(lr){1-2}\cmidrule(lr){3-6}\cmidrule(lr){7-10}
$Y$ & $X$ & $A$ & $B$ & $C$  & $\sigma_{Y|X}$  & $A$ & $B$ & $C = C_{self}$  & $\sigma_{Y|X}$ \\
\midrule

$L_{\rm X}$ & $T$ & $0.91^{+0.16}_{-0.15}$ & $2.89^{+0.14}_{-0.13}$ & $1.59^{+0.86}_{-0.93}$ & $0.78^{+0.08}_{-0.07}$ & $0.89^{+0.16}_{-0.15}$ & $2.93^{+0.12}_{-0.12}$ & 1 & $0.80^{+0.07}_{-0.07}$\\ 
$L_{\rm X}$ & $M_{\rm gas}$ & $0.89^{+0.02}_{-0.02}$ & $1.10^{+0.03}_{-0.02}$ & $1.44^{+0.25}_{-0.26}$ & $0.30^{+0.02}_{-0.02}$ & $0.88^{+0.02}_{-0.02}$ & $1.07^{+0.02}_{-0.02}$ & 2 & $0.30^{+0.02}_{-0.02}$\\ 
$L_{\rm X}$ & $Y_{\rm X}$ & $1.20^{+0.04}_{-0.04}$ & $0.83^{+0.02}_{-0.02}$ & $1.50^{+0.33}_{-0.35}$ & $0.29^{+0.03}_{-0.03}$ & $1.20^{+0.04}_{-0.04}$ & $0.83^{+0.02}_{-0.02}$ & 8/5 & $0.29^{+0.03}_{-0.03}$\\ 
\midrule 
$L_{\rm bol}$ & $T$ & $1.02^{+0.14}_{-0.14}$ & $3.01^{+0.13}_{-0.12}$ & $2.69^{+0.74}_{-0.78}$ & $0.70^{+0.07}_{-0.06}$ & $0.96^{+0.15}_{-0.14}$ & $3.13^{+0.12}_{-0.12}$ & 1 & $0.76^{+0.07}_{-0.06}$\\ 
$L_{\rm bol}$ & $M_{\rm gas}$ & $0.86^{+0.03}_{-0.03}$ & $1.19^{+0.03}_{-0.03}$ & $1.86^{+0.29}_{-0.30}$ & $0.32^{+0.02}_{-0.02}$ & $0.86^{+0.02}_{-0.02}$ & $1.16^{+0.02}_{-0.02}$ & 7/3 & $0.31^{+0.02}_{-0.02}$\\ 
$L_{\rm bol}$ & $Y_{\rm X}$ & $1.12^{+0.03}_{-0.03}$ & $0.90^{+0.02}_{-0.02}$ & $1.83^{+0.27}_{-0.28}$ & $0.28^{+0.02}_{-0.02}$ & $1.12^{+0.03}_{-0.03}$ & $0.90^{+0.02}_{-0.02}$ & 9/5 & $0.28^{+0.02}_{-0.02}$\\ 
\midrule 
$M_{\rm gas}$ & $T$ & $0.83^{+0.13}_{-0.13}$ & $2.41^{+0.11}_{-0.11}$ & $0.21^{+0.74}_{-0.79}$ & $0.67^{+0.07}_{-0.06}$ & $0.77^{+0.13}_{-0.12}$ & $2.47^{+0.11}_{-0.10}$ & -1 & $0.72^{+0.06}_{-0.06}$\\

\bottomrule
\end{tabular}
\label{tab:scalingresults}
\tablefoot{Fitted relation is of the form $Y = A~Y_{\rm piv}  \left (\frac{X}{X_{\rm piv}} \right )^B \left (\frac{E(z)}{E(z_{\rm piv})} \right )^C$ with a log-normal intrinsic scatter $\sigma_{\rm Y|X}$ (in natural log). Pivot values of the observables are provided in Table~\ref{tab:pivots}. Each relation is fitted twice; first leaving the redshift evolution exponent ($C$) free, and second with a redshift evolution exponent fixed to the corresponding self-similar value (see Table~\ref{tab:selfsimilar} for the self-similar exponents). Details of the modeling and fitting the scaling relations can be found in Sect.~\ref{sec:modelingandfitting}. Errors are 1$\sigma$ uncertainties calculated from the second half of the MCMC chains.}
\end{table*}
\section{Results}
\label{sec:results}

Scaling relations between X-ray observables are tools for understanding the ICM physics for various mass scales and evolution of the ICM with redshift, while the relations between observables and cluster mass are used for facilitating cosmology with cluster number counts. In this section we examine the $L-T,\ L-M_{\rm gas},\ L-Y_{\rm X}$, and $\ M_{\rm gas}-T$ scaling relations, using both $L_{\rm X}$ and $L_{\rm bol}$, and provide extensive comparisons with the literature. Owing to the high soft-band sensitivity of the eROSITA, we were able to include a large number of low-mass, low-luminosity clusters in our study, down to the soft band luminosities of $8.64\times10^{40}$~ergs~s$^{-1}$ and masses ($M_{500}$) of $6.86 \times 10^{12}\,$M$_{\odot}$. In the eFEDS field alone, we detect a total of 68 low-mass groups with $M_{500}<10^{14}$~M$_{\odot}$ that are fully included in our analysis. eROSITA will be revolutionary in both ICM studies and cosmology in this regard as it will extend cluster samples to much lower luminosities and lower masses than ever reached before. We first describe our method and lay the groundwork with the eFEDS sample with this work, and will push the mass and luminosity limits down with our ongoing work on the eRASS1 sample. One other important aspect is the fact that the eROSITA group and cluster samples are uniformly selected and the selection function is well understood with the help of our full-sky eROSITA simulations. 

There are several complications in comparing scaling relation results in the literature with our results. These are linked to the form of the fitted scaling relations, the energy band of the extracted observables, and the assumed cosmology, and the instrument calibration also varies from one study to another. To overcome these difficulties, we apply corrections before we compare them with our results. In these comparison plots, we use the self-similar redshift evolution as the common reference point and convert the observables accordingly. The standard energy band we use in this paper for the extraction of observables is the $0.5-2.0$~keV band. To convert normalizations of scaling relations involving luminosities obtained in the $0.1-2.4$~keV energy band ($L_{0.1-2.4}$), we faked an unabsorbed {\sc APEC} spectrum within {\tt XSPEC} and calculated a conversion factor of 1.64 for a cluster that has a temperature of 3.26~keV, an abundance of 0.3, and a redshift of 0.33. These redshift and temperature values are the median values of our sample (see Table~\ref{tab:pivots}). Changing the temperatures and redshifts affects the conversion factor by a few percent, which is consistent with the findings of \citet{Lovisari2020}. We therefore applied the same conversion factor to all other works using the $0.1-2.4$~keV energy band. Lastly, we convert the relations assuming a dimensionless Hubble constant of 0.6774 which is the value we use in this work. The corrections are only applied to the normalizations, and therefore the slopes and redshift evolution exponents of previously reported relations remain unchanged.
\begin{figure*}
\centering
\begin{tabular}{c}
\includegraphics[width=0.32\textwidth]{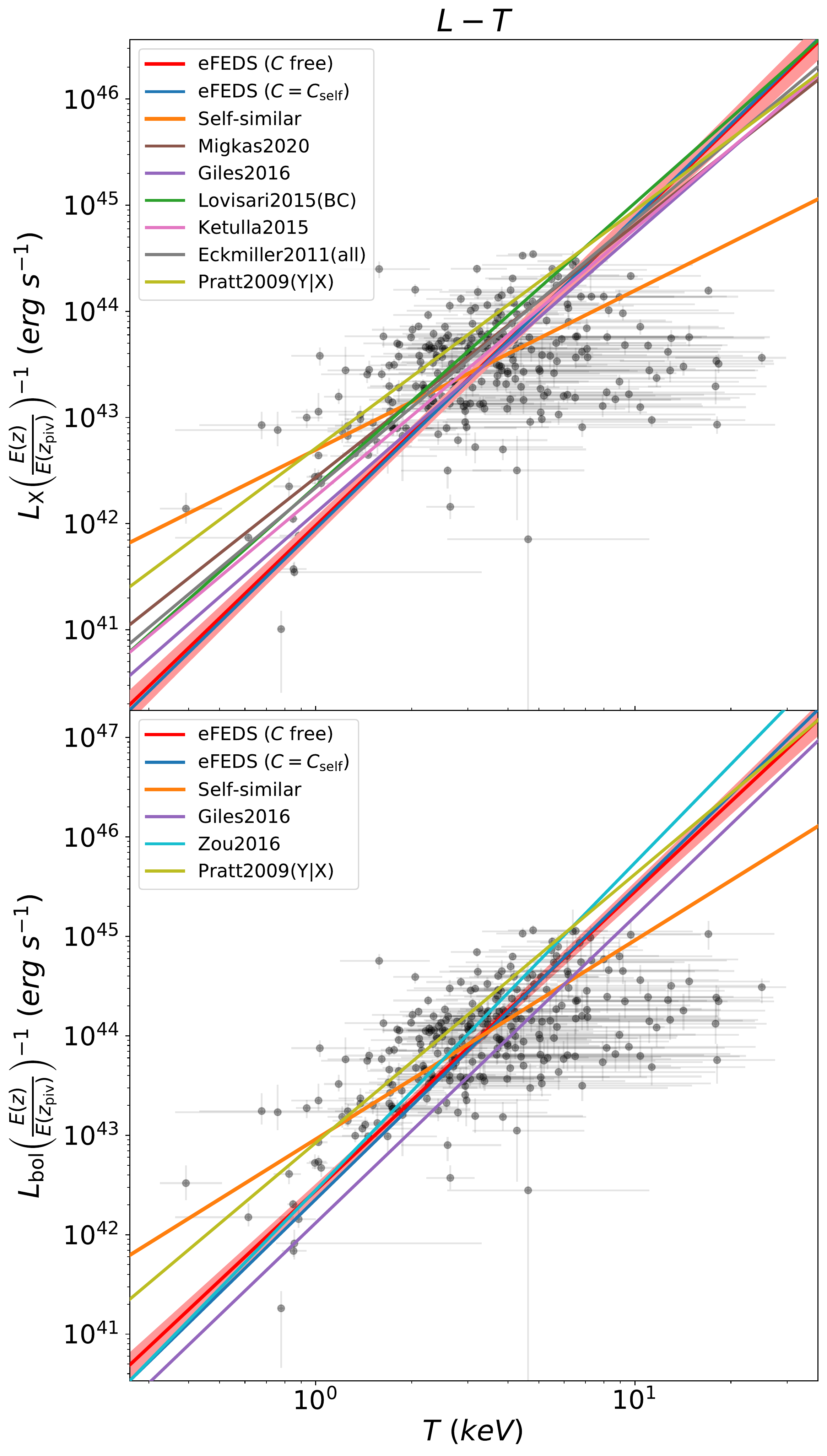} 
\includegraphics[width=0.32\textwidth]{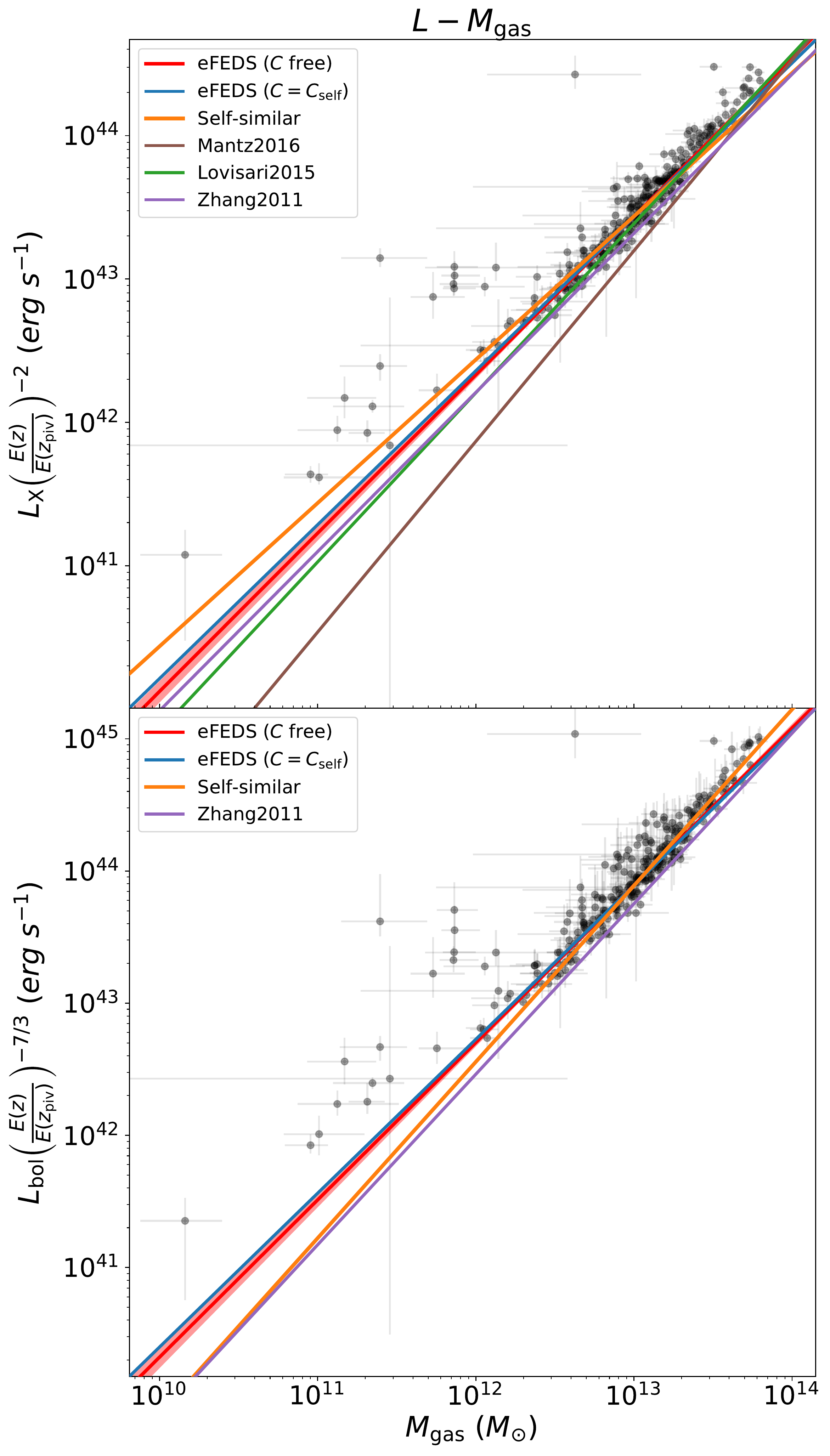} 
\includegraphics[width=0.32\textwidth]{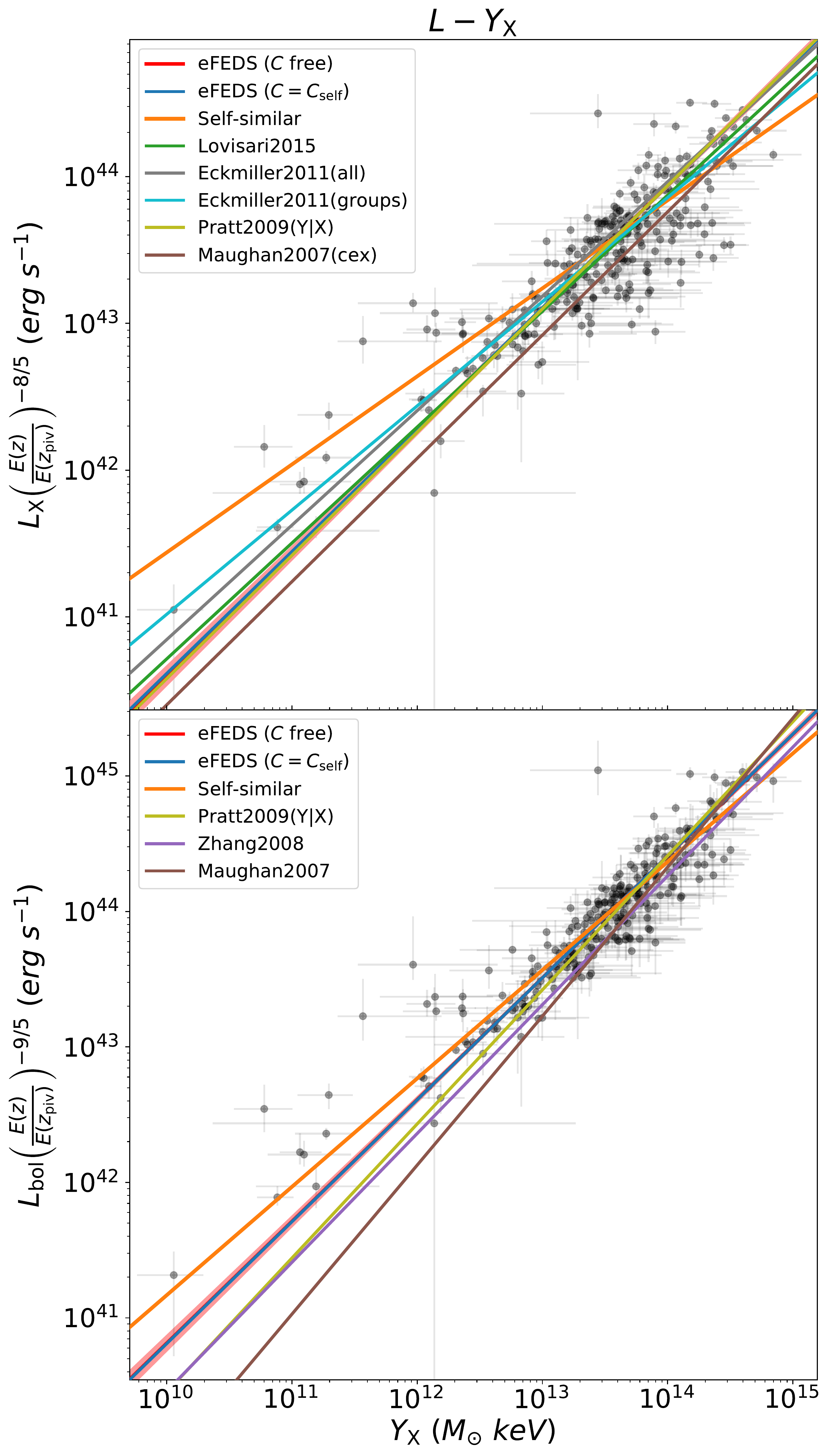} \\
\includegraphics[width=0.32\textwidth]{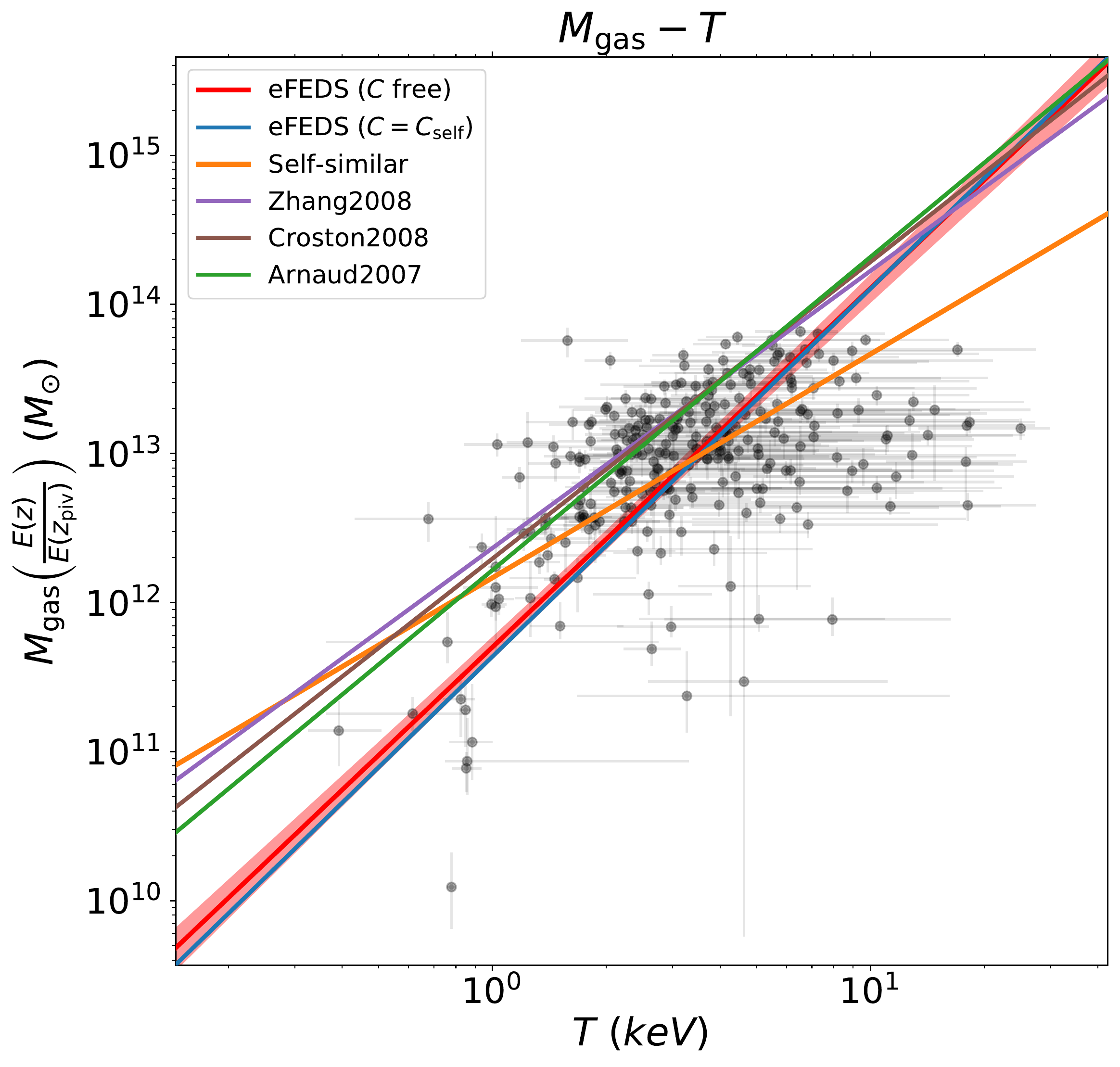} 
\end{tabular}
\caption{Comparison between our best-fit $L-T$, $L-M_{\rm gas}$,  $L-Y_{\rm X}$, and $M_{\rm gas}-T$ relations, the self-similar model and other studies in the literature \citep{Maughan+07,arnaud2007, Zhang+08, Pratt2009, Eckmiller2011, zhang2011,Lovisari2015, Mantz2016}. Grey circles are the eFEDS clusters. In order to achieve consistency, a cosmology and energy band conversion is applied to the previously reported results (see Sect.~\ref{sec:results} for the details).
\label{fig:comp}}
\end{figure*}
\label{lmgascompararison}

Another challenge in comparisons of scaling relations involving the ICM temperature is the calibration differences between various X-ray telescopes. It has been shown that calibration differences between  {\it Chandra} and  XMM{\it-Newton} are dependent  on the energy band and can be as large as a factor of two for hot clusters with temperatures $>10$~keV in the soft band ($0.7-2$~keV)  \citep{Schellenberger2015}. However, this difference is small, namely of 10\%--15\% in the full $0.7-7$~keV band for low-temperature clusters ($<4$~keV) to which we are sensitive in the eFEDS observations. Our preliminary calibration studies with eROSITA showed that, in general, eROSITA temperatures are in good agreement with {\it Chandra} and XMM{\it-Newton} temperatures \citep{Sanders2021, Veronica2021, Iljenkarevic2021, Whelan2021}. \citet{Turner2021} recently cross-matched the eFEDS cluster catalog \citep{liu2021a} with the XMM{\it-Newton} Cluster Survey \citep[XCS,][]{Romer2001} sample and found luminosities of 29 cross-matched clusters to be in excellent agreement. They also compared the temperatures of 8 clusters that are measured with both telescopes and found XMM measurements to have slightly higher temperatures on average. In order to better understand the instrumental differences, more extensive studies should be performed with a cluster sample containing a larger range of temperatures using the survey data. This will be further investigated in future eROSITA projects.

\begin{figure*}
\centering
\begin{tabular}{c}
\includegraphics[width=\textwidth]{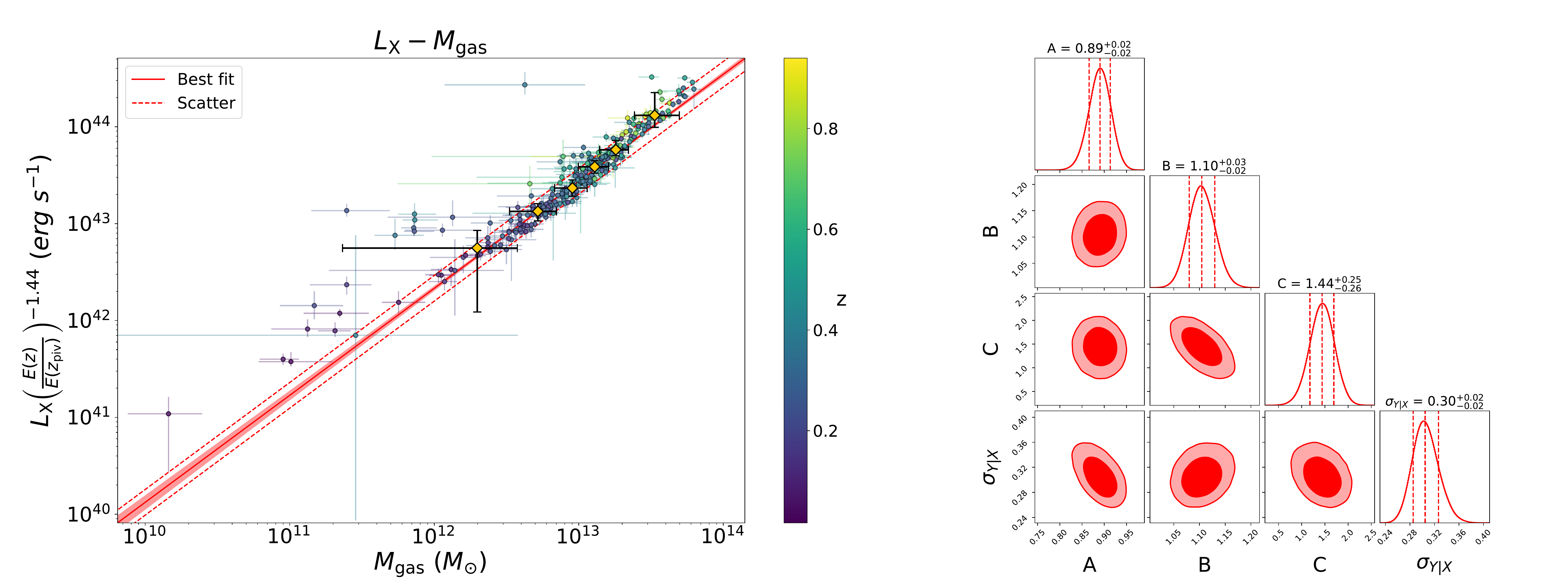} \\
\includegraphics[width=\textwidth]{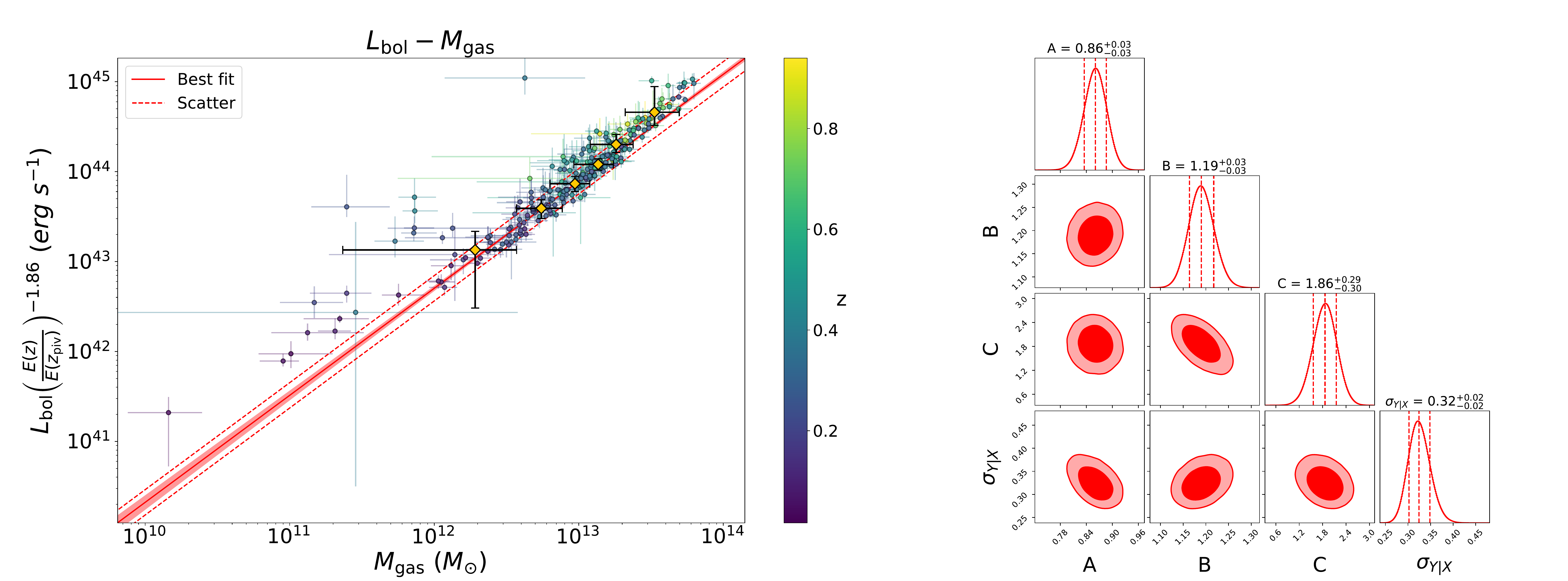}
\end{tabular}
\caption{$L-M_{\rm gas}$ scaling relations and the posterior distributions of the scaling relations parameters. {\it Left:} Soft band ($0.5-2.0$ keV) X-ray luminosity ($L_{\rm X}$), bolometric ($0.01-100$ keV) luminosity  ($L_{\rm bol}$), gas mass ($M_{\rm gas}$), and redshift ($z$) measurements of the $\mathcal{L}_{\rm det} > 15$, $\mathcal{L}_{\rm ext} > 15$ sample and the best-fit scaling relation models. The light-red shaded area indicates the 1$\sigma$ uncertainty of the mean of the log-normal model (see Eq.~\ref{eqn:lognormalmodel}) and the dashed red line indicates the best-fit standard deviation ($\sigma_{L|M_{\rm gas}}$) around the mean. Orange diamonds indicate median gas mass measurements obtained from clusters between luminosity quantiles. {\it Right:} Parameter constraints of the $L_{\rm X}-M_{\rm gas}$ and $L_{\rm bol}-M_{\rm gas}$ relations obtained from the second half of the MCMC chains. Marginalized posterior distributions are shown on the diagonal plots and the joint posterior distributions are shown on off-diagonal plots. Red dashed vertical lines indicate the 32th, 50th, and 68th percentiles and contours indicate 68\% and 95\% credibility regions.
\label{figLMgas}}
\end{figure*}
\subsection{$L_{\rm X}-T$ and $L_{\rm bol}-T$ relations}
The two main observables from X-rays, luminosity, and temperature reflect different but complexly related features of the ICM in clusters. On one hand, luminosity is proportional to the square of the electron density, and therefore it is highly sensitive to the distribution of the hot gas. On the other hand, the temperature is related to the average kinetic energy of the baryons in the ICM. Both luminosity and temperature are subject to gravitational and nongravitational effects in a different manner and this makes their relation nontrivial \citep[see][for a more detailed discussion]{Giodini2013}. Hence, a better understanding of the $L-T$ relation will shed light on the history of the heating and cooling mechanisms of clusters.

In the self-similar scenario \citep{kaiser1986}, the relation between luminosity, temperature, and redshift follows
\begin{equation}
L_{\rm X} \propto T^{3/2} E(z)
\end{equation}
and
\begin{equation}
L_{\rm bol} \propto T^{2} E(z).
\end{equation}
However, a plethora of studies report steeper $L_{\rm X}-T$ ($B\sim2.5$) and $L_{\rm bol}-T$ ($B\sim3.0$) relations compared to their self-similar predictions \citep[e.g.,][]{Pratt2009,Eckmiller2011, Maughan2012,Hilton2012,Kettula2015,Lovisari2015,Zou2016,Giles2016,Molham2020}. 

Our best-fit results for the $L_{\rm X}-T$  relation are presented in Table~\ref{tab:scalingresults} where we report a slope of $B=2.89^{+0.14}_{-0.13}$, a redshift evolution dependence of $C=1.59^{+0.86}_{-0.93}$, and a scatter of $\sigma_{\rm L_{\rm X}|T}=0.78^{+0.08}_{-0.07}$. The best-fit model is shown in Fig.~\ref{figLT}. In general, our results agree well with studies that account for the selection biases. A comparison of our results with some others can be found in Fig.~\ref{fig:comp}. Our best-fit slope is significantly steeper at a $\sim 11\sigma$ confidence level than the self-similar expectation ($B_{\rm self}=3/2$). Our relation is slightly steeper than the slopes reported for the XXL sample, $B=2.63\pm0.15$ \citep{Giles2016}, and the combined Northern ROSAT All-Sky Survey (NORAS) plus ROSAT-ESO Flux Limited X-ray Survey (REFLEX) samples, $B=2.67\pm0.11$ \citep{Lovisari2015}, but all three agree well within $1.3\sigma$. We note that these latter authors fully account for selection effects in their analysis and both of these latter samples are the most similar to the eFEDS sample because they also contain a significant fraction of low-mass clusters. Our slope is also slightly steeper than the slopes found in \citet{Eckmiller2011} ($B=2.52\pm0.17$) and \citet{Kettula2015} ($B=2.52\pm0.17$) but is consistent with both within $1.7\sigma$. The slope of $B=2.24\pm0.25$ reported in \citet{Pratt2009} is $2.3\sigma$ shallower than our results. Our best-fit redshift evolution for the $L_{\rm X}-T$ relation is in agreement with the self-similar scenario ($C_{\rm self}=1$) and with \citet{Giles2016} within $1\sigma$. Furthermore, our redshift evolution also agrees well with most of the other results because of the large error bars which indicate the redshift evolution could not be constrained as well as other parameters. When we fix the evolution term to the self-similar value, we find a steeper slope of ($B=2.93\pm0.12$). This slope is $\sim4\sigma$ away from the slope reported in \citet{Migkas2020} ($B=2.38\pm0.08$) if their temperature correction is taken into account. Our best-fit intrinsic scatter for the $L_{\rm X}-T$ relation agrees very well with \citet{Pratt2009} ($\sigma_{\rm L_{\rm X}|T}=0.76\pm0.14$) whereas it is in $~3\sigma$ tension with \citet{Giles2016} ($\sigma_{\rm L_{\rm X}|T}=0.47\pm0.07$). \citet{Lovisari2015} ($\sigma_{\rm L_{\rm X}|T}=0.56$) and \citet{Eckmiller2011} ($\sigma_{\rm L_{\rm X}|T}=0.63$) also reported slightly smaller intrinsic scatter results compared to our findings, but a statistical comparison cannot be made because of the lack of error bars in their scatter measurements.

For the $L_{\rm bol}-T$ relation, we find a slope of $B=3.01^{+0.13}_{-0.12}$, a redshift evolution term of $C=2.69^{+0.74}_{-0.78}$, and a scatter of $\sigma_{\rm L_{\rm bol}|T}=0.70^{+0.07}_{-0.06}$. Both the slope and the redshift evolution are steeper than the self-similar expectation of $B_{\rm self}=2$ at a $8.5\sigma$ level and $C_{\rm self}=1$ at a $~2\sigma$ level. Due to the temperature dependence of the X-ray emissivity, the $L-T$ scaling relation involving the bolometric luminosity is expected to be steeper than that of the soft-band luminosity for the same cluster by a factor of $\propto n_{e}^{2}T^{0.5}$. The slope in this case agrees very well with \citet{Giles2016} ($B=3.08\pm0.15$), \citet{Zou2016} ($B=3.29\pm0.33$), and \citet{Pratt2009} ($B=2.70\pm0.24$). When we fix the redshift evolution to the self-similar value, we obtain a steeper slope of $B=3.13\pm0.12$. Our best-fit slope with fixed redshift evolution is also consistent with the slopes in \citet{Giles2016} and \citet{Zou2016} within uncertainties whereas \citet{Migkas2020} found a slope that is shallower by $4.5\sigma$  ($B=2.46\pm0.09$). \citet{Maughan2012} ($B=3.63\pm0.27$) on the other hand, found a steeper slope than both the self-similar model and our results when they included the cluster cores. \citet{Maughan2012} reported that when they limit their sample to relaxed cool core clusters, they find a much shallower slope of $B=2.44\pm 0.43$ indicating that the discrepancy observed here could be due to their samples being heavily affected by the selection effects which we take into account by using realistic simulations in our analysis. The intrinsic scatter of the $L_{\rm bol}-T$ relation is lower compared to the best-fit value of our $L_{\rm X}-T$ relation, but they agree within the error bars. \citet{Pratt2009} reported $\sigma_{\rm L_{\rm bol}|T}=0.73\pm0.14$, which is consistent with our results for the $L_{\rm bol}-T$ relation within uncertainties. Our best-fit intrinsic scatter is slightly higher  than the findings reported in \citet{Zou2016} ($\sigma_{\rm L_{\rm bol}|T}=0.47\pm0.11$) and \citet{Giles2016} ($\sigma_{\rm L_{\rm bol}|T}=0.47\pm0.07$), but within $1.8$ and $2.5\sigma$ statistical uncertainty, respectively.

\subsection{$L_{\rm X}-M_{\rm gas}$ and $L_{\rm bol}-M_{\rm gas}$ relations}

Luminosity and gas mass are two tightly related observables because of their mutual dependence on electron density, and therefore a strong correlation is expected
between them. Measurement of their correlation whilst taking into account selection effects and the mass function with a large sample allows us to test the theorized relation between these observables. According to the self-similar model, they are connected as
\begin{equation}
L_{\rm X} \propto M_{\rm gas} E(z)^{2}
\end{equation}
and
\begin{equation}
L_{\rm bol} \propto M_{\rm gas}^{4/3} E(z)^{7/3}.
\end{equation}

Our best-fit results for the $L_{\rm X}-M_{\rm gas}$ and $L_{\rm bol}-M_{\rm gas}$ relations are provided in Table~\ref{tab:scalingresults} and in Fig.~\ref{figLMgas}. A comparisons of these results with previous work is shown in Fig.~\ref{fig:comp}. We report a slope of $B=1.10^{+0.03}_{-0.02}$, a redshift evolution term of $C=1.44^{+0.25}_{-0.26}$, and a scatter of $\sigma_{\rm L_{\rm X}|M_{\rm gas}}=0.30\pm0.02$. The slope is in tension with the self-similar expectation at a $5\sigma$ level, but the redshift evolution is consistent with the self-similar model within $2\sigma$ confidence for the $L_{\rm X}-M_{\rm gas}$ relation. When we fix the redshift evolution to the self-similar value, the slope does not change significantly ($B=1.07\pm0.02$). \citet{zhang2011} obtained a slope of $B=1.11\pm0.03$ from the 62 clusters in the HIFLUGCS sample which is consistent with our measurements. Their slope for the cool-core clusters ($B=1.09\pm0.05$) is similar to what they found for their whole cluster sample, but the best-fit slope for their noncool-core clusters is steeper ($B=1.20\pm0.06$).
\citet{Lovisari2015} studied the scaling properties of a complete X-ray-selected galaxy group sample and found the slope of  the $L_{\rm X}-M_{\rm gas}$ relation for galaxy groups to be $B=1.02\pm0.24$, which is slightly shallower than but still consistent with the result they obtained for more massive clusters, $B=1.18\pm0.07$. Both of these measurements are consistent with our slope. On the other hand, a flux-limited sample of 139 clusters compiled from the ROSAT All-Sky Survey catalog has a steeper slope with $B=1.34\pm0.05$ for the $L_{\rm X}-M_{\rm gas}$ relation \citep{Mantz2016}. The result of these latter authors is more than $4\sigma$ away from our measurement. This discrepancy might be due to the fact that the \citet{Mantz2016} sample is dominated by massive luminous clusters (their lowest luminosity system is about as bright as our most luminous systems), while the eFEDS sample is composed of low-mass clusters and groups. There are not many studies in the literature reporting intrinsic scatter of the $L_{\rm X}-M_{\rm gas}$ relation. Therefore, we were only able to compare our results with those of \citet{zhang2011}, who found $\sigma_{\rm L_{\rm X}|M_{\rm gas}}=0.14\pm2,$ which is significantly lower ($5.5\sigma$) than our results.

On the other hand, we find the best-fit parameters of the slope, the evolution term, and the scatter of the $L_{\rm bol}-M_{\rm gas}$ relation are $B=1.19\pm0.03$, $C=1.86^{+0.29}_{-0.30}$, and $\sigma_{L_{\rm bol}|M_{\rm gas}}=0.32\pm0.02$, respectively. Similarly, the slope is $\sim5\sigma$ away from the self-similar model, while the redshift evolution is fully consistent with the model. This relation has received much less attention in the literature. \citet{zhang2011} found a slope of $B=1.29\pm0.05$ when they fitted their whole sample. Their reported slope is less steep for cool-core clusters ($B=1.24\pm0.05$) relative to the noncool-core clusters ($B=1.42\pm0.06$). The slope of the whole sample is fully consistent with our measurements within $2\sigma$. Similar to the $L_{\rm X}-M_{\rm gas}$ relation, we could only compare our best-fit intrinsic scatter for the $L_{\rm bol}-M_{\rm gas}$ relation with the results of \citet{zhang2011}, who report $\sigma_{\rm L_{\rm bol}|M_{\rm gas}}=0.21\pm2,$ which is in $4\sigma$ tension with our results.

One additional point is that there is a noticeable deviation around the gas mass of $\sim10^{12}~$M$_{\odot}$ in Fig.~\ref{figLMgas}. The low-mass groups tend to show higher luminosity relative to the mass determined from the scaling relations. The slope and normalization of this power-law relation are mostly governed by the higher mass clusters. The low-mass groups would prefer a shallower $L_{\rm X}-M_{\rm gas}$ power-law slope relative to the high-mass clusters. This observed trend is fully consistent with the $L_{\rm X}-M_{\rm gas}$ relation reported by \citet{Lovisari2015}, who similarly observed the relation getting shallower at the group scale but within error bars.

\begin{figure*}
\centering
\begin{tabular}{c}
\includegraphics[width=\textwidth]{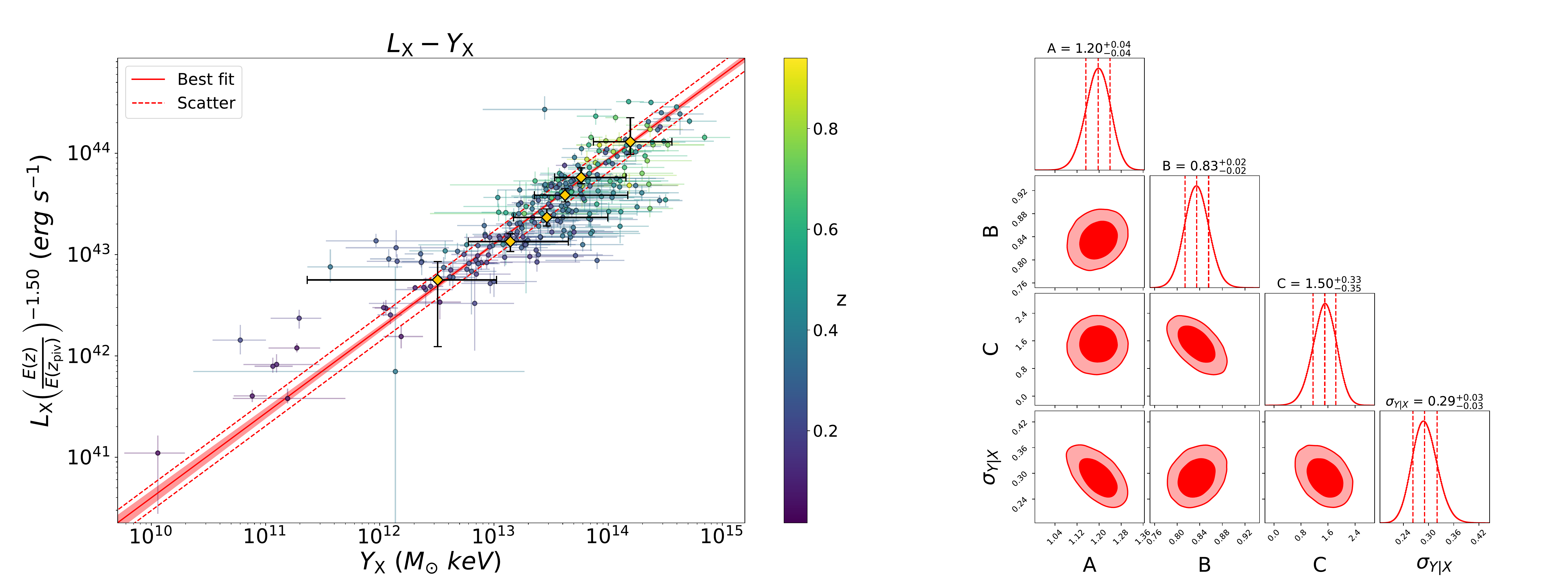} \\
\includegraphics[width=\textwidth]{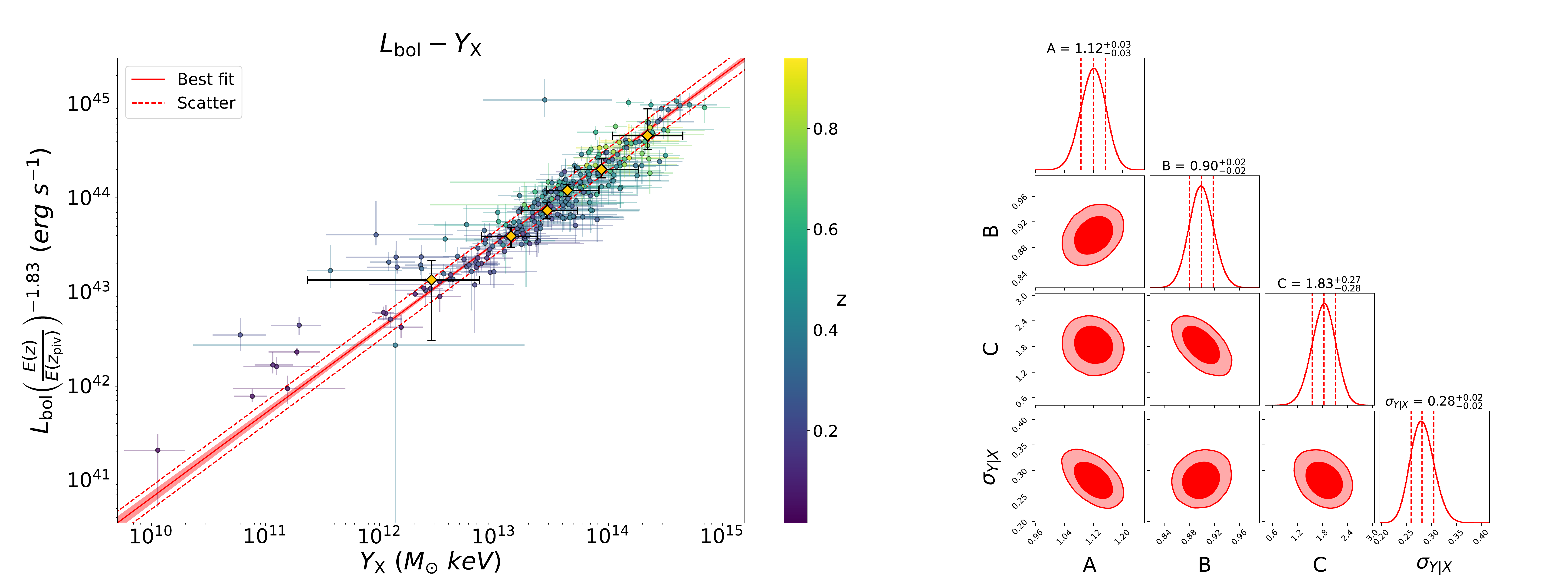}
\end{tabular}
\caption{$L-Y_{\rm X}$ scaling relations and the posterior distributions of the scaling relations parameters. {\it Left:} Soft band ($0.5-2.0$ keV) X-ray luminosity ($L_{\rm X}$), bolometric ($0.01-100$ keV) luminosity  ($L_{\rm bol}$), $Y_{\rm X}$ , and redshift ($z$) measurements of the $\mathcal{L}_{\rm det} > 15$, $\mathcal{L}_{\rm ext} > 15$ sample and the best-fit scaling relation models. The light-red shaded area indicates the 1$\sigma$ uncertainty of the mean of the log-normal model (see Eq.~\ref{eqn:lognormalmodel}) and the dashed red line indicates the best-fit standard deviation ($\sigma_{L|Y_{\rm X}}$) around the mean. Orange diamonds indicate median $Y_{\rm X}$ measurements obtained from clusters between luminosity quantiles. {\it Right:} Parameter constraints of the $L_{\rm X}-Y_{\rm X}$ and $L_{\rm bol}-Y_{\rm X}$ relations obtained from the second half of the MCMC chains. Marginalized posterior distributions are shown on the diagonal plots and the joint posterior distributions are shown on off-diagonal plots. Red dashed vertical lines indicate the 32th, 50th, and 68th percentiles and contours indicate 68\% and 95\% credibility regions.
\label{figLY}}
\end{figure*}
\subsection{$L_{\rm X}-Y_{\rm X}$ and $L_{\rm bol}-Y_{\rm X}$ relations}
The accurate mass indicator, $Y_{\rm X}$, first introduced by \citet{Kravtsov2006}, shows a low intrinsic scatter with mass and has a tight relation with the Sunyaev Zel'dovich (SZ) effect observable Compton-$y$ parameter, $Y_{\rm SZ}$ \citep[e.g.,][]{Maughan+07, Benson2013, Mantz2016, Bulbul2019, Andrade-Santos2021}. Because of this strong correlation, scaling relations involving $Y_{\rm X}$ can be used as a connection in multi-wavelength studies of galaxy clusters. Numerical simulations suggest that nongravitational effects have a small influence on this mass proxy compared to other X-ray observables \citep{Nagai2007}.

According to the self-similar model, luminosity is expected to depend on $Y_{\rm X}$ and redshift as
\begin{equation}
L_{\rm X} \propto Y_{\rm X}^{3/5} E(z)^{8/5}
\end{equation}
and
\begin{equation}
L_{\rm bol} \propto Y_{\rm X}^{4/5} E(z)^{9/5}.
\end{equation}

Our results for the best-fit $L_{\rm X}-Y_{\rm X}$ relations are listed in Table~\ref{tab:scalingresults} and  plotted in Fig.~\ref{figLY}, while a comparison with the literature is provided in Fig. \ref{fig:comp}. We find a slope of $B=0.83\pm0.02,$  a redshift evolution exponent of $C=1.50^{+0.33}_{-0.35}$, and an intrinsic scatter of $\sigma_{L_{\rm X}|Y_{\rm X}}=0.29\pm0.03$ for the $L_{\rm X}-Y_{\rm X}$ scaling relation. Our slope for the $L_{\rm X}-Y_{\rm X}$ relation is $11.5\sigma$ steeper than that predicted by the self-similar model. The redshift evolution of the $L_{\rm X}-Y_{\rm X}$ relation is slightly shallower than the self-similar expectation but is consistent within the uncertainties. Our slope is consistent with the results presented in \citet{Maughan+07} ($B=0.84\pm0.03$) and with that of \citet{Lovisari2015} ($B=0.79\pm0.03$). Our results for the same relation are within $1.8\sigma$ statistical uncertainty of the HIFLUGCS+groups sample of \citet{Eckmiller2011} and within $2.2\sigma$ of their groups-only sample. These latter authors find slopes of $B=0.78\pm 0.02$ and $B=0.71\pm0.05$ for the HIFLUGCS+groups and  groups only samples, respectively, where the latter is within $\sim2\sigma$ from the self-similar expectation. Our best-fit intrinsic scatter is in good agreement ($~1.5\sigma$) with the findings of \citet{Pratt2009} ($\sigma_{L_{\rm X}|Y_{\rm X}}=0.41\pm0.07$). \citet{Eckmiller2011} ($\sigma_{L_{\rm X}|Y_{\rm X}}=0.46$) and \citet{Lovisari2015} ($\sigma_{L_{\rm X}|Y_{\rm X}}=0.51$) report higher values for the intrinsic scatter of the $L_{\rm X}-Y_{\rm X}$ relation, but these measurements are presented without error bars and therefore a statistical comparison with our findings cannot be made.

For the $L_{\rm bol}-Y_{\rm X}$ relation, we find a slope of $B=0.90\pm0.02$, a redshift evolution exponent of $C=1.83^{+0.27}_{-0.28}$, and an intrinsic scatter of $\sigma_{L_{\rm bol}|Y_{\rm X}}=0.28\pm0.02$. The slope shows a $5\sigma$ deviation from self-similarity. \citet{Maughan+07} find an even larger deviation from the self-similarity, measuring a slope of $B=1.10\pm0.04$. Also, in \citet{Zhang+08} and \citet{Pratt2009}, the authors reported steeper slopes of $B=0.95\pm0.08$ and $B=1.04\pm0.06$ where the former agrees well with our results within statistical uncertainties whereas the latter is $2.2\sigma$ higher. Numerical simulations show a similar scenario. \citet{Biffi+14} reports this slope to be $B=0.94\pm0.02$, which is also slightly steeper than our results and significantly steeper than the self-similar value. Our redshift evolution for the $L_{\rm bol}-Y_{\rm X}$ relation is consistent with the self-similar prediction within the uncertainties. A similar redshift evolution was measured in \citet{Maughan+07}, with $C=2.2\pm0.1$ which is $<1.5\sigma$ away from our finding. Our best-fit intrinsic scatter for the $L_{\rm bol}-Y_{\rm X}$ relation is slightly smaller ($1.5\sigma$) compared to the value reported in \citet{Pratt2009} ($\sigma_{L_{\rm bol}|Y_{\rm X}}=0.38\pm0.06$). \citet{Maughan+07} reported a similar value ($\sigma_{L_{\rm bol}|Y_{\rm X}}=0.36\pm0.03$) for the intrinsic scatter that is in $2.2\sigma$ tension with our best-fit value.

\begin{figure*}
\centering
\begin{tabular}{c}
\includegraphics[width=\textwidth]{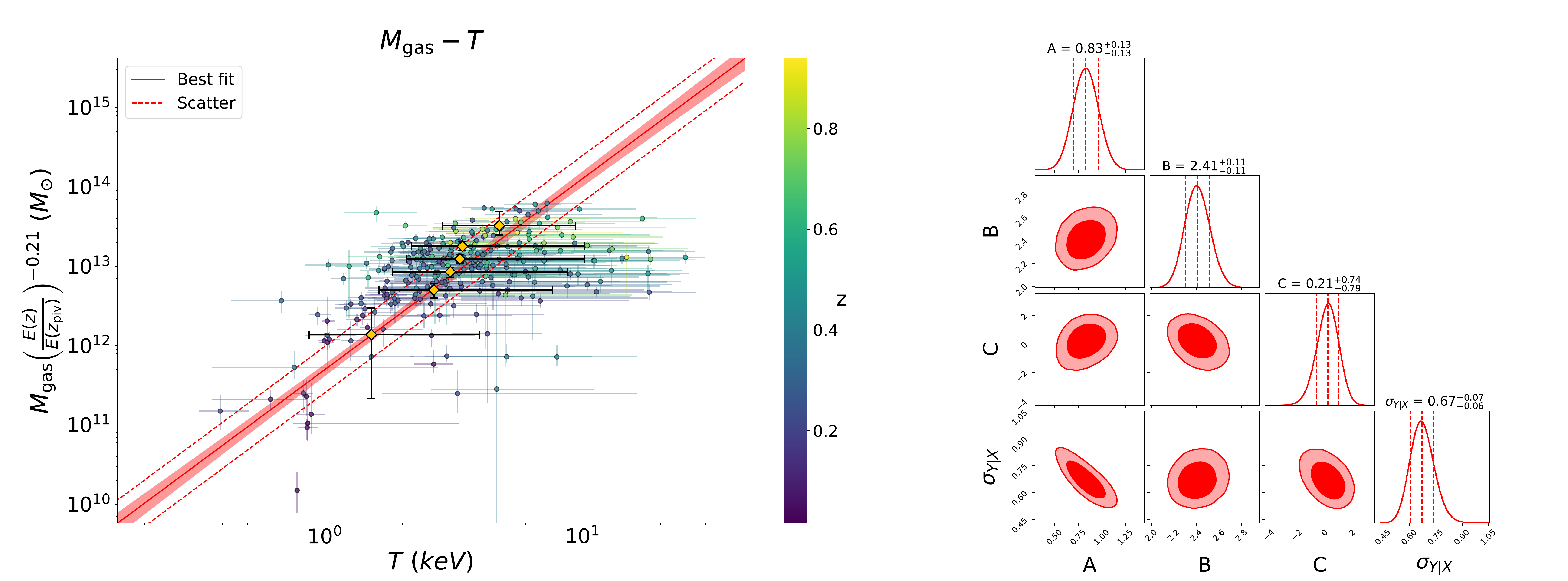}
\end{tabular}
\caption{$M_{\rm gas}-T$ scaling relation and the posterior distributions of the scaling relation parameters. {\it Left:} Gas mass ($M_{\rm gas}$), temperature ($T$), and redshift ($z$) measurements of the $\mathcal{L}_{\rm det} > 15$, $\mathcal{L}_{\rm ext} > 15$ sample and the best-fit scaling relation models. The light-red shaded area indicates the 1$\sigma$ uncertainty of the mean of the log-normal model (see Eq.~\ref{eqn:lognormalmodel}) and the dashed red line indicates the best-fit standard deviation ($\sigma_{M_{\rm gas}|T}$) around the mean. Orange diamonds indicate median temperature measurements obtained from clusters between gas mass quantiles. {\it Right:} Parameter constraints of the $M_{\rm gas}-T$ relation obtained from the second half of the MCMC chains. Marginalized posterior distributions are shown on the diagonal plots and the joint posterior distributions are shown on off-diagonal plots. Red dashed vertical lines indicate 32th, 50th, and 68th percentiles and contours indicate 68\% and 95\% credibility regions.
\label{figMgasT}}
\end{figure*}
\subsection{$M_{\rm gas}-T$ relation}

$M_{\rm gas}-T$ and $L_{\rm X}-T$ relations are conjugates of each other because of the tight correlation between $M_{\rm gas}$ and $L_{\rm X}$. However, we still expect to see differences as $M_{\rm gas}$ has a linear dependence on electron density whereas $L$ has a quadratic dependence. We fit the $M_{\rm gas}-T$ relation following the similar framework as in the sections above with minor changes. We use the $L_{\rm X}$ flavored selection function by converting $M_{\rm gas}$ to $L_{\rm X}$ because we do not have a selection function involving $M_{\rm gas}$ from simulations. This one-to-one conversion in principle should not introduce a large bias to our results because $L_{\rm X}$ and $M_{\rm gas}$ are tightly correlated and the scatter between them is relatively low. 

Based on the self-similar model, gas mass and temperature should be related to each other via
\begin{equation}
M_{\rm gas} \propto T^{3/2} E(z)^{-1}.
\end{equation}

Our results for the $M_{\rm gas}-T$ relation are listed in Table~\ref{tab:scalingresults} and shown in Fig.~\ref{figMgasT}. We obtain a slope of $B=2.41\pm0.11$, a redshift evolution exponent of $C=0.21^{+0.74}_{-0.79}$, and a scatter of $\sigma_{M_{\rm gas}|T}=0.67^{+0.07}_{-0.06}$. Our slope is $8.3\sigma$ steeper than the self-similar model. We find a positive redshift evolution which is expected to be negative in the self-similar case, but our result agrees with the self-similar prediction within $1.5\sigma$ statistical uncertainty. A comparison of our results with the literature is given in Fig.~\ref{fig:comp}.
In general, the slope of the $M_{\rm gas}-T$ relation reported in the literature is close to $\sim1.9$, which is steeper than the self-similar expectation. Reported slopes in the literature show a dependency on the mass range of the parent sample. For instance, \citet{arnaud2007} reports a slope of $2.10\pm0.05$ based on the XMM{\it-Newton} observations of ten relaxed nearby clusters. Consistently, \citet{croston2008} found $1.99\pm0.11$ using the 31 clusters in the REXCESS sample and \citet{Zhang+08} obtained $1.86\pm0.19$ with XMM{\it-Newton} data for 37 LoCuSS clusters. These slopes are shallower than the results reported here, with a $~2.5\sigma$ difference. The discrepancy could be due to the different selection of the samples compared here.

We find a factor of approximately two difference when we compare our best-fit normalization results with those of \citet{arnaud2007} and \citet{croston2008} at their pivot temperature value (5 keV). To investigate this difference and test our results, we reconstructed the $M_{\rm gas}-T$ relation using our best-fit $L_{\rm X}-T$ and $L_{\rm X}-M_{\rm gas}$ relations, which are in agreement with the most recent studies in the literature taking into account the selection effects. We obtain a relation of
\begin{equation}\label{eqn:derived_mt}
M_{\rm gas} = 1.02~M_{\rm gas,piv}  \left (\frac{\rm T}{T_{\rm piv}} \right )^{2.63} \left (\frac{E(z)}{E(z_{\rm piv})} \right )^{0.14}.
\end{equation}
\noindent The normalization, slope, and evolution terms are $<2\sigma$ away from the best-fit $M_{\rm gas}-T$ relation which indicates that our results for the $M_{\rm gas}-T$ relation would be in good agreement with the previous results if the selection effects were taken into account. We argue that the observed discrepancy arises due to the combined effect of two main differences between our analyses and the other analyses reported in the literature for the same relation. The first difference is that we include selection effects in our work and therefore measure a steeper slope for the $M_{\rm gas}-T$ relation compared to the previously reported results. Our steeper slope agrees well with previous findings because $M_{\rm gas}$ is a very good $L_{\rm X}$ proxy and many studies, including ours, show that the best-fit slope of the $L_{\rm X}-T$ relation is found to be steeper when the selection effects are taken into account. The second difference is that our sample includes a larger fraction of low-mass clusters compared to the other samples. If the cluster populations were similar, we would not observe such a difference in normalization even if the slopes did not match. Therefore, in our case, the populations and the slopes being different combine and result in the observed mismatch. Additionally, using a converted flavored selection function might have also contributed to the discrepancy, but its effect is expected to be much smaller because the relation between $L_{\rm X}$ and $M_{\rm gas}$ is tight and the scatter is low.

\section{Discussion}

Slopes of the scaling relations between X-ray observables studied in this work show deviations from the self-similar model by $4-11.5\sigma$  confidence levels. These deviations are often attributed to the departures from the assumptions in the self-similar \citep{kaiser1986} model in the literature. We discuss two potential reasons for the observed discrepancy in the eFEDS sample in this section. 

The most commonly proposed explanations in the literature for the departures from self-similarity challenge two major assumptions in the model. First, clusters are assumed to have a spherically symmetric gas distribution that is in hydrostatic equilibrium. Secondly, physical processes are majorly driven by the gravitational force, and the other effects are negligible in shaping the observed physical state of clusters. Observational data and numerical simulations indicate that both of these assumptions may not hold, and this can lead to departures from the self-similar expectation. Nongravitational processes such as AGN feedback, galactic winds, and star formation introduce extra energy to the system, heat the gas, and increase the entropy in the core \citep[e.g.,][]{Voit2005, Walker2012, Bulbul2016}. AGN feedback in particular can play an important role in shaping the gas physics,  especially in low-mass clusters and groups that dominate the eFEDS extended source sample. AGN activity can expel gas to the outskirts for lower mass haloes because of their shallower potentials wells. As the larger fraction of gas is removed from the centers of low-mass haloes, their luminosity decreases \citep[e.g.,][]{Puchwein2008}. The most massive clusters with deeper potential wells, higher total mass, and the ICM temperature are less affected by the nongravitational effects. When the clusters and groups are fit together, the lower luminosity of the groups and low-mass clusters leads to a steeper slope of the $L_{\rm X}-T$ scaling relations when the cores are included. This result is consistent with numerical simulations \citep{Puchwein2008, Schaye2010, Biffi+14, Truong2018, Henden2018, Henden2019}, and the observational data in the literature \citep{Eckmiller2011, Maughan2012, Pratt2009, Zou2016, Giles2016, Migkas2020, Lovisari2021}. Another proposed reason for these steep slopes is the use of the overdensity radii $R_{500}$ derived from the X-ray masses calculated assuming the hydrostatic equilibrium \citep[see ][ for discussion]{Lovisari2021}. If the radius $R_{500}$ is biased low because of the unaccounted-for nonthermal pressure in the ICM, the luminosity extracted in this radius would be lower. The temperatures are less impacted by this effect, because of large uncertainties. However, in this work,  we use the overdensity radii $R_{500}$ obtained from the HSC weak lensing measurements uniformly for low-mass groups and clusters in the eFEDS sample \citep{Chiu2021}. We argue that the (mass-dependent) hydrostatic equilibrium bias and radius of extraction do not have a major effect in this work and cannot explain the steepening slope of the $L_{\rm X}-T$ scaling relations. The $M_{\rm gas}-T$ relations should be affected by the AGN feedback similarly but less severely than the $L_{\rm X}-T$ relation because of the linear dependence of $M_{\rm gas}$ on the number density of electrons, i.e., $M_{\rm gas}\propto n_{e}$ while $L_{\rm X}\propto n_{e}^{2}$ through X-ray emissivity. The expected steepening in $M_{\rm gas}-T$ should be less prominent as a result of this effect. This is fully consistent with our results assuming that the discrepancy is attributed to nongravitational effects. Another important issue in comparing various results in the literature is the calibration differences between X-ray telescopes. The number density and luminosity measurements are expected to be consistent between X-ray telescopes, namely {\it Chandra} and XMM{\it-Newton} \citep{Bulbul2019}; however, significant band-dependent disagreements have been reported for cluster ICM temperature measurements \citep{Schellenberger2015}. Given that the reported discrepancies between X-ray instruments are small in the soft X-ray band where the temperature measurements of most of our clusters lie, we do not expect that the slope differences are due to these calibration effects. 

For $L_{\rm X}-Y_{\rm X}$ and $L_{\rm X}-M_{\rm gas}$ scaling relations, the effect of Malmquist bias, i.e., preferential sampling of bright objects, can clearly be seen and is often prominent in X-ray-selected samples. We note that these biases and selection effects are fully accounted for in our selection function, and therefore should not bias our best-fit scaling relations. We observe mild deviations from the self-similar model on both scaling relations in low $M_{gas}$ and $Y_{\rm X}$ regimes. The mass proxy $Y_{\rm X}$ shows low intrinsic scatter with cluster mass in the literature \citep{Kravtsov2006, Eckmiller2011, Bulbul2019}. As the ICM temperature scales with total mass, we expect to see a similar trend with low-scatter in the $L_{\rm X}-Y_{\rm X}$ scaling relations. Indeed, the $L_{\rm X}-Y_{\rm X}$ scaling relations show a tight correlation for massive clusters. Along the lines of what we observe, as the group scales dominate the sample, the intrinsic scatter becomes larger. We find that the departures from the self-similarity are significant for both of the relations which is consistent with the results reported in the literature and numerical simulations \citep{Eckmiller2011, Biffi+14, Lovisari2015}.

The self-similar model predicts cosmology-dependent redshift evolution for the scaling relations between observables and cluster mass. This dependence is introduced through the overdensity radius (and the critical density), which is inversely related to the evolution of the Hubble parameter with redshift z, $E(z)=H(z)/H_{0}$.
We do not find significant departures from the self-similar evolution with redshift in any of our relations. All show self-similar redshift evolution $<2.5\sigma$ confidence level. There are contradictory
reports in the literature as to the evolution of scaling relations. Some studies report self-similar redshift evolution with redshift \citep{Giodini2013}, while some studies report no evolution \citep{Maughan+07, Pacaud2007}. Larger samples, covering a wide redshift range, will be available with the eRASS data, which can be used to constrain the redshift evolution of scaling relations and test the self-similar model.

In this work, we investigate the scaling relations between X-ray observables of the clusters of galaxies and galaxy groups by fully modeling the selection effects. Our method of obtaining the selection function relies on realistic simulations of the eROSITA observations. This is the most robust way of modeling the selection effects as long as the simulated population of sources is representative of the observed sample. The advantage of this method lies in the fact that the same detection and reduction methods are applied to both simulated observations and the eROSITA data self-consistently \citep{Clerc2018,2020Comparat}. In these simulations, cluster surface brightness profiles are created by making use of the previously measured profiles of cluster and group samples that span a wide range of mass and redshift; they use X-COP, SPT, XXL, and HIFLUGCS cluster samples. This method allows the profiles to be consistent with the observations, except in the low-L, low-z regime where we probe a mass and redshift space that is poorly explored by previous X-ray observations. This led to a slight excess in the number of detected simulated clusters by the pipeline in this parameter space, the presence of which is barely visible in Fig.~\ref{Ldists}. The mild difference does not have any effect on our best fitting relations because our likelihood takes the redshift ($z$) and detection ($I$) information of clusters as given, $\mathcal{L}(\hat{Y}_{\rm all},\hat{X}_{\rm all}|I, \theta,z)$, such that in our analysis, the shape of $P(I|L_{\rm X},z)$ as a function of $L_{\rm X}$ is more important than the relative normalizations at different redshifts, $ P(I|z)= \int P(I|L_{\rm X},z) P(L_{\rm X}|z) dL_{\rm X}$.

Following our analysis of the eFEDS observations, this less populated mass--luminosity range will be filled with eFEDS clusters, and therefore surface brightness profiles of simulated clusters at these regimes will be improved for modeling the selection function for the future eRASS observations. Proper modeling of the selection effects will be particularly important for placing constraints on cosmological parameters using eROSITA observations \citep{Clerc2018}.

We also test our method by comparing the model-predicted number of detected clusters ($N_{\rm det}$) and the number of clusters in our observed sample ($\hat{N}_{\rm det}$) as also presented in \citet{Giles2016}. However, we find that comparing the predicted and observed cluster numbers is not informative because the predictions have large uncertainties driven mostly by the propagated errors from our scaling relation analysis and the weak-lensing mass-calibrated scaling relation analysis \citep{Chiu2021}. To give an example, we compare the observed number of detected clusters for the $\mathcal{L}_{\rm det} > 15$, $\mathcal{L}_{\rm ext} > 15$ sample (265) with the model-predicted number for the $L_{\rm X}-T$ relation similar to the \citet{Giles2016}. We calculated $N_{\rm det}$ as
\begin{equation}\label{eqn:ndet}
N_{\rm det} = \int \int \int_{L_{\rm X}, T, z} P(I|L_{\rm X},z) P (L_{\rm X} | T,\theta_{L_{\rm X}T}, z) \frac{dn}{dT} \frac{dV_{\rm c}}{dz} dL_{\rm X} dT dz
,\end{equation}
where $ \frac{dV_{\rm c}}{dz}$ is the differential comoving volume shell spanning a solid angle of $\Omega_{\rm eFEDS} = 140/(180/\pi)^2$, $\frac{dn}{dT}$ is the temperature function calculated as described in Sect.~\ref{sec:likelihood}, and $\theta_{L_{\rm X}T}$ is the best-fit parameters of the $L_{\rm X}-T$ relation. While calculating $N_{\rm det}$, first we only propagate the errors of the best-fit parameters ($\theta_{L_{\rm X}T}$) using MCMC chains and a pivot value of $M_{\rm piv}=1.4\times10^{14} M_{\odot}$, which is the median of the eFEDS sample. We find the model-predicted number of detected clusters to be $N_{\rm det, L_{\rm X}T}=301.2^{+42.5}_{-49.4}$. When we further propagate both the uncertainties of $\theta_{L_{\rm X}T}$ and the best-fit weak-lensing mass-calibrated scaling relation parameters ($A_\mathcal{X}$, $B_\mathcal{X}$, $\gamma_\mathcal{X}$), we find the new measurement to be $N_{\rm det, L_{\rm X}T}=309.3^{+134.2}_{-86.1}$. In this case, the observed number of clusters is consistent with the predicted number of clusters within the uncertainties. The difference in the absolute values might be due to the selection function, or the cosmology-dependent normalization of the mass function. In this work and simulations, we use the Planck cosmology \citep{Planck2016} with the \citet{tinker08} mass function. Larger samples of clusters of galaxies will soon be available through the eROSITA All-Sky observations and these will provide sufficient statistics to constrain the cosmology simultaneously with the scaling relations \citep[see][for the cosmology forecast]{Pillepich2012}.

Decreasing the scatter is of significant importance in reducing the systematic error on the constraints of cosmological parameters through cluster counts. Cool-core and relaxed clusters are reported to show less scatter on the scaling relations relative to the mergers \citep{Vikhlinin2009a, Mantz2010a}. The dynamical states of the eFEDS clusters and groups were presented by \citet{Ghirardini2021}. The dynamically relaxed clusters compose  30\% to 40\% of this sample, and therefore using them reduces the statistical power of our measurements. Additionally, the use of the relaxed cluster in the scaling-relation fits requires a selection function characterized in terms of these morphological parameters and a  dynamical-state-dependent mass function \citep[e.g.,][]{Seppi2021}. This form of selection function is not available yet. We therefore leave this work to future studies of the eRASS data.

\section{Conclusions}
The eFEDS is of a similar depth to the final eROSITA All-Sky Survey in Equatorial regions. In this field, we detect 542 galaxy clusters and groups in the extent-selected sample with an addition of 347 clusters of galaxies in the point source samples \citep{liu2021a, Klein2021, Bulbul2021}. In this work, we present the X-ray properties ($L_{\rm X}$, $L_{\rm bol}$, $T$, $M_{\rm gas}$, $Y_{\rm X}$) of the all eFEDS clusters and groups measured in two apertures; core-included ($r<R_{500}$) and core-excluded ($0.15R_{500}<r<R_{500}$). The overdensity radius $R_{500}$ is obtained from the HSC weak-lensing mass estimates provided by \citet{Chiu2021}, allowing our measurements to be free of bias from the hydrostatic equilibrium assumption. This work clearly demonstrates that the cluster ICM emission will be significantly detected for most of the clusters in the mass and redshift ranges out to $R_{500}$ at this depth.

Additionally, we provide the $L-T,\ L-M_{\rm gas},\ L-Y_{\rm X}$, and $M_{\rm gas}-T$ scaling relations between these (core-included) X-ray observables for a subsample of clusters. We only consider the extent-selected sample, where we can characterize the selection effects using the state-of-the art simulations. Contamination of the main eFEDS clusters and groups sample by AGNs and false detections due to fluctuations  is
on the order of 20\% \citep[see ][for details]{liu2021a, LiuT2021}. To reduce the contamination of the sample to under 10\%, we further apply the cuts on the extent and detection likelihoods of $\mathcal{L}_{\rm det} > 15$ and $\mathcal{L}_{\rm ext} > 15$. We note that, apart from the $\mathcal{L}_{\rm det}$ and $\mathcal{L}_{\rm ext}$ cuts, we have not applied any further cleaning to the sample, such as optical cross-matching. These cuts reduce the sample size to 265 clusters and groups spanning a redshift range of $0.017<z<0.94$, a soft-band luminosity range of $8.64 \times 10^{40}~{\rm erg}~{\rm s}^{-1}<L_{\rm X}<3.96 \times 10^{44}~{\rm erg}~{\rm s}^{-1}$, and a mass range of $6.86 \times10^{12}~$M$_{\odot}<M_{500}< 7.79\times10^{14}$~M$_{\odot}$. In this sample, we find 68 low-mass galaxy groups with $<10^{14}$M$_{\odot}$, which are uniformly selected with the rest of the massive clusters in the sample. We investigated these seven relations by taking into account both the selection effects and the cosmological distributions of observables. Fitting was performed twice for each of the seven relations, first with a redshift evolution exponent left free and the second with a redshift evolution exponent fixed to the corresponding self-similar value. The main conclusions of our analysis are as follows:

\begin{enumerate}
\item[$\boldsymbol{-}$] The eFEDS scaling-relation results between X-ray observables in general are in good agreement with the recently reported results. However, we find significant tension with the self-similar expectation for all scaling relations studied here. We suggest that these deviations indicate that the nongravitational effects such as the feedback mechanisms play a key, nonnegligible role in shaping the observed physical properties of the clusters, especially in the low-mass group regime. 
Specifically, the scaling-relation results we present in this paper for the $L-T$ relation agree well with the results from the samples that are similar to the eFEDS sample when the selection function is taken into account \citep{Giles2016, Lovisari2015, Eckmiller2011}. Our results for the $L-T$ relation also agree well with the {\sc fable} and {\sc macsis} simulations where they include baryonic feedback \citep{Puchwein2008, Henden2019, Biffi+14, Barnes2017}.

\item[$\boldsymbol{-}$] The largest scatter we measure is for the $L_{\rm X}-T$ and $L_{\rm bol}-T$ relations. We find $\sigma_{L_{\rm X}|T}=0.80\pm0.07$, $\sigma_{ L_{\rm bol}|T}=0.76^{+0.07}_{-0.06}$ when we fix the redshift evolution to the self-similar value and $\sigma_{L_{\rm X}|T}=0.78^{+0.08}_{-0.07}$, $\sigma_{ L_{\rm bol}|T}=0.70^{+0.07}_{-0.06}$ when the evolution is let free. This intrinsic scatter is mostly driven by the groups. The lowest scatter is measured for the $L_{\rm X}-Y_{\rm X}$ and $L_{\rm bol}-Y_{\rm X}$ relations with $\sigma_{ L_{\rm X(bol)}|Y_{\rm X}}=0.29(0.28)\pm0.03(0.02)$. This result shows that in addition to $Y_{\rm X}$ being a good mass indicator, it is also a good proxy for the X-ray properties of the ICM.

\item[$\boldsymbol{-}$] The redshift evolution of the scaling relations of the seven scaling relations we examined is broadly consistent with the self-similar model. Fixing the redshift evolution exponent to the corresponding self-similar value only changes the best-fit slopes by $<1\sigma$ from its previous value obtained with a free exponent. Larger samples of clusters and groups are required for constraining the evolution of these relations with redshift.

\item[$\boldsymbol{-}$] We find that  the $M_{\rm gas}-T$ relation differs from the previous results in the literature by a factor of approximately two in normalization. This could be driven by the lack of proper consideration of the selection effects in the previously reported results or by the fact that the eFEDS sample contains a much greater number of low-mass clusters and groups than the compared samples. This difference might  also partially be due to the lack of a selection function with the $M_{\rm gas}$ flavor. Inclusion of X-ray observables other than $L_{\rm X}$ and $L_{\rm bol}$ in the simulations is an ongoing project, and will help to understand the effects of such phenomena.
\end{enumerate}

This work extends the study of X-ray scaling relations to a sample that is dominated by low-mass clusters and groups. It crates a further avenue by which to study ICM physics for a new population of low-mass clusters and groups, as well as massive clusters at a wide redshift range.
These initial results demonstrate the capability of eROSITA to detect the ICM emission out to $R_{500}$ for a large number of clusters detected at the final depth of the All Sky Survey. We note that this depth will be exceeded by the first All Sky Survey for a limited area, allowing the observables to be measured out to $R_{500}$ or beyond for a subsample of clusters. These measurements will provide access to increasingly stringiest constraints on the mass and redshift evolution of the scaling relations. eFEDS observations not only allow us to put tight constraints on the scaling relation parameters but also allow us to test our selection function, which will be used in future statistical studies with eROSITA. 

\begin{acknowledgement}

This work is based on data from eROSITA, the soft X-ray instrument aboard SRG, a joint Russian-German science mission supported by the Russian Space Agency (Roskosmos), in the interests of the Russian Academy of Sciences represented by its Space Research Institute (IKI), and the Deutsches Zentrum f{\"{u}}r Luft und Raumfahrt (DLR). The SRG spacecraft was built by Lavochkin Association (NPOL) and its subcontractors, and is operated by NPOL with support from the Max Planck Institute for Extraterrestrial Physics (MPE).

The development and construction of the eROSITA X-ray instrument was led by MPE, with contributions from the Dr. Karl Remeis Observatory Bamberg \& ECAP (FAU Erlangen-Nuernberg), the University of Hamburg Observatory, the Leibniz Institute for Astrophysics Potsdam (AIP), and the Institute for Astronomy and Astrophysics of the University of T{\"{u}}bingen, with the support of DLR and the Max Planck Society. The Argelander Institute for Astronomy of the University of Bonn and the Ludwig Maximilians Universit{\"{a}}t Munich also participated in the science preparation for eROSITA. The eROSITA data shown here were processed using the eSASS software system developed by the German eROSITA consortium.

The Hyper Suprime-Cam (HSC) collaboration includes the astronomical communities of Japan and Taiwan, and Princeton University. The HSC instrumentation and software were developed by the National Astronomical Observatory of Japan (NAOJ), the Kavli Institute for the Physics and Mathematics of the Universe (Kavli IPMU), the University of Tokyo, the High Energy Accelerator Research Organization (KEK), the Academia Sinica Institute for Astronomy and Astrophysics in Taiwan (ASIAA), and Princeton University. Funding was contributed by the FIRST program from the Japanese Cabinet Office, the Ministry of Education, Culture, Sports, Science and Technology (MEXT), the Japan Society for the Promotion of Science (JSPS), Japan Science and Technology Agency (JST), the Toray Science Foundation, NAOJ, Kavli IPMU, KEK, ASIAA, and Princeton University. 

This paper makes use of software developed for the Large Synoptic Survey Telescope. We thank the LSST Project for making their code available as free software at  http://dm.lsst.org

This paper is based [in part] on data collected at the Subaru Telescope and retrieved from the HSC data archive system, which is operated by the Subaru Telescope and Astronomy Data Center (ADC) at National Astronomical Observatory of Japan. Data analysis was in part carried out with the cooperation of Center for Computational Astrophysics (CfCA), National Astronomical Observatory of Japan. The Subaru Telescope is honored and grateful for the opportunity of observing the Universe from Maunakea, which has the cultural, historical and natural significance in Hawaii.

E.B. acknowledges financial support from the European Research Council (ERC) Consolidator Grant under the European Union’s Horizon 2020 research and innovation programme (grant agreement No 101002585). N.C. is supported by CNES.

\end{acknowledgement}

\bibliographystyle{aa}
\bibliography{xray_scaling}

\onecolumn

\newpage
\onecolumn
\fontsize{7.0pt}{0cm}\selectfont

\begin{landscape}
\begin{longtable}{lccccccccccccccccccc}
\caption{\label{tab:R500} X-ray observables of eFEDS clusters measured within $R_{500}$ and between $0.15R_{500}-R_{500}$.}\\
\toprule\midrule
Cluster & ID\_SRC & RA & Dec & $\mathcal{L}_{\rm ext}$ & $\mathcal{L}_{\rm det}$ & $z$ & $R_{\rm 500}$ & $T$ & $L_{\rm X}$ & $L_{\rm bol}$ & $M_{\rm gas}$ & $Y_{\rm X}$ & $T_{\rm cex}$ & $L_{\rm X,cex}$ & $L_{\rm bol, cex}$ & $t_{\rm exp}$ \\
(eFEDS~J+) &  & (deg) & (deg) &  &  &  & (arcmin) & (keV) & ($10^{42}$ erg s$^{-1}$) & ($10^{42}$ erg s$^{-1}$) & ($10^{12}$ M$_{\odot}$) & ($10^{12}$ M$_{\odot}$ keV) & (keV) & ($10^{42}$ erg s$^{-1}$) & ($10^{42}$ erg s$^{-1}$) & (s) \\

\midrule
\endfirsthead
\caption{continued.}\\
\toprule\midrule
Cluster & ID\_SRC & RA & Dec & $\mathcal{L}_{\rm ext}$ & $\mathcal{L}_{\rm det}$ & $z$ & $R_{\rm 500}$ & $T$ & $L_{\rm X}$ & $L_{\rm bol}$ & $M_{\rm gas}$ & $Y_{\rm X}$ & $T_{\rm cex}$ & $L_{\rm X,cex}$ & $L_{\rm bol, cex}$ & $t_{\rm exp}$ \\
(eFEDS~J+) &  & (deg) & (deg) &  &  &  & (arcmin) & (keV) & ($10^{42}$ erg s$^{-1}$) & ($10^{42}$ erg s$^{-1}$) & ($10^{12}$ M$_{\odot}$) & ($10^{12}$ M$_{\odot}$ keV) & (keV) & ($10^{42}$ erg s$^{-1}$) & ($10^{42}$ erg s$^{-1}$) & (s) \\
\midrule
\endhead
\midrule
\endfoot
\bottomrule
\endlastfoot

082626.5-003429 & 28993 & 126.6108 & -0.5748 & 8.5 & 5.0 & 0.161 & 2.404 & ${5.1}_{-3.4}^{+13.4}$ & $<5.7$ & $<30$ & $<2.5$ & $<36$ & ${4.1}_{-2.7}^{+12.9}$ & $<4.9$ & $<24$ & 184\\ 
082751.7-002853 & 11248 & 126.9655 & -0.4816 & 12.8 & 27.9 & 0.257 & 2.620 & ${1.80}_{-0.63}^{+2.11}$ & $<12$ & $<33$ & $<5.0$ & $<15$ & ${1.73}_{-0.63}^{+9.17}$ & $<7.0$ & $<21$ & 707\\ 
082808.7-001003 & 4800 & 127.0366 & -0.1677 & 28.5 & 62.5 & 0.076 & 5.487 & ${0.994}_{-0.058}^{+0.096}$ & ${2.40}_{-0.34}^{+0.48}$ & ${4.58}_{-0.62}^{+0.99}$ & ${1.12}_{-0.20}^{+0.27}$ & ${1.13}_{-0.22}^{+0.31}$ & ${1.08}_{-0.14}^{+0.26}$ & ${1.62}_{-0.34}^{+0.52}$ & ${3.22}_{-0.70}^{+1.12}$ & 675\\ 
082820.5-000721 & 4169 & 127.0856 & -0.1228 & 42.4 & 81.4 & 0.845 & 1.894 & ${5.5}_{-2.0}^{+5.1}$ & ${273}_{-45}^{+43}$ & ${990}_{-180}^{+330}$ & ${42.5}_{-6.5}^{+6.2}$ & ${233}_{-84}^{+218}$ & ${4.9}_{-2.2}^{+5.7}$ & ${232}_{-44}^{+49}$ & ${790}_{-170}^{+310}$ & 748\\ 
082840.6-000500 & 7991 & 127.1692 & -0.0836 & 18.4 & 37.5 & 0.320 & 2.504 & ${8.7}_{-4.9}^{+11.2}$ & ${14.7}_{-3.0}^{+2.7}$ & ${65}_{-22}^{+44}$ & ${5.7}_{-1.7}^{+1.8}$ & $<190$ & ${7.0}_{-3.8}^{+11.4}$ & ${8.3}_{-2.2}^{+3.9}$ & ${35}_{-13}^{+26}$ & 856\\ 
082859.9+010756 & 14973 & 127.2500 & +1.1323 & 19.2 & 14.4 & 0.354 & 2.522 & ${6.9}_{-3.8}^{+13.3}$ & ${27.4}_{-6.8}^{+7.8}$ & ${115}_{-41}^{+91}$ & ${11.9}_{-2.7}^{+3.2}$ & $<420$ & ${5.7}_{-3.1}^{+10.8}$ & ${25.9}_{-6.8}^{+6.9}$ & ${95}_{-31}^{+69}$ & 464\\ 
082952.7+002139 & 7528 & 127.4697 & +0.3611 & 7.7 & 38.6 & 0.420 & 1.874 & ${6.3}_{-4.0}^{+14.4}$ & ${14.3}_{-2.8}^{+4.7}$ & ${58}_{-21}^{+49}$ & ${3.7}_{-1.4}^{+1.6}$ & $<130$ & ${6.0}_{-3.4}^{+13.6}$ & ${7.5}_{-2.6}^{+4.3}$ & ${31}_{-14}^{+27}$ & 1099\\ 
082955.4+004131 & 3810 & 127.4810 & +0.6922 & 24.2 & 80.3 & 0.940 & 1.555 & ${5.6}_{-2.6}^{+8.7}$ & ${217}_{-39}^{+47}$ & ${790}_{-160}^{+410}$ & ${28.8}_{-4.5}^{+4.6}$ & ${152}_{-66}^{+269}$ & ${4.1}_{-1.4}^{+4.1}$ & ${180}_{-41}^{+45}$ & ${580}_{-140}^{+160}$ & 1020\\
\end{longtable}
\tablefoot{Electronic version of the table is available at the CDS. Column 1: cluster name. Column 2: unique source ID presented in the eFEDS source catalog \citep{Brunner2021}. Columns 3 and 4: RA and Dec. Columns 5 and 6: extent and detection likelihoods. Column 7: redshift. Column 8: $R_{500}$ estimates calculated from the  $M_{500}$ measurements presented in \citep{Chiu2021}. Column 9: temperature measured within $R_{500}$. Column 10: soft band (0.5--2~keV) luminosity measured within $R_{500}$. Column 11: bolometric (0.01--100~keV) luminosity measured within $R_{500}$. Column 12: gas mass measured within $R_{500}$. Column 13: X-ray analog of integrated Compton-$y$ parameter measured within $R_{500}$. Column 14: core-excised temperature measured between $0.15R_{500}-R_{500}$. Column 15: soft band (0.5--2~keV) core-excised luminosity measured between $0.15R_{500}-R_{500}$. Column 16:  bolometric (0.01--100~keV) core-excised luminosity measured between $0.15R_{500}-R_{500}$. Column 17: unvignetted exposure time measured at the X-ray center of the cluster. X-ray observable measurements $<2\sigma$ are presented as $2\sigma$ upper limits except $T$ and $T_{\rm cex}$.}
\end{landscape}
\newpage

\clearpage

\LTcapwidth=\textwidth
\newpage
\onecolumn
\fontsize{10.0pt}{0cm}\selectfont
\begin{longtable}{ l c c c c c c }
\caption{\label{tab:sb_fit} X-ray best fitting parameters for \citet{Vikhlinin+06} model: $n_e^2(r) = n_0^2 \left( \frac{r}{r_s} \right)^{-\alpha} \left( 1 + \left( \frac{r}{r_s} \right)^2 \right)^{-3\beta+\alpha/2} \left(  1 + \left( \frac{r}{r_s} \right)^3 \right)^{-\epsilon/3}$.}\\
\toprule\midrule
Cluster & ID\_SRC & $n_0^2$ & $r_s$ & $\epsilon$ & $\beta$ & $\alpha$ \\
(eFEDS~J+) & & (10$^{-7}$ cm$^{-6}$) & (arcsec) & & & \\

\midrule
\endfirsthead
\caption{continued.}\\
\toprule\midrule
Cluster & ID\_SRC & $n_0^2$ & $r_s$ & $\epsilon$ & $\beta$ & $\alpha$ \\
(eFEDS~J+) & & (10$^{-7}$ cm$^{-6}$) & (arcsec) & & & \\
\midrule
\endhead
\midrule
\endfoot
\bottomrule
\endlastfoot

082626.5-003429 & 28993 & ${4.0}_{-3.0}^{+36.4}$ & ${72}_{-52}^{+65}$ & ${2.5}_{-1.7}^{+1.7}$ & ${1.56}_{-0.77}^{+0.66}$ & ${1.09}_{-0.71}^{+1.01}$ \\
082751.7-002853 & 11248 & ${1.9}_{-1.6}^{+32.7}$ & ${72}_{-46}^{+67}$ & ${2.6}_{-1.7}^{+1.6}$ & ${1.59}_{-0.77}^{+0.64}$ & ${1.54}_{-0.98}^{+1.17}$ \\
082808.7-001003 & 4800 & ${0.75}_{-0.50}^{+3.53}$ & ${224}_{-82}^{+87}$ & ${1.5}_{-1.0}^{+1.8}$ & ${1.16}_{-0.57}^{+0.90}$ & ${1.42}_{-0.71}^{+0.34}$ \\
082820.5-000721 & 4169 & ${8.9}_{-7.4}^{+31.8}$ & ${125}_{-45}^{+72}$ & ${3.0}_{-1.8}^{+1.5}$ & ${1.62}_{-0.87}^{+0.67}$ & ${1.12}_{-0.74}^{+0.53}$ \\
082840.6-000500 & 7991 & ${1.7}_{-1.4}^{+12.1}$ & ${123}_{-47}^{+60}$ & ${2.6}_{-1.8}^{+1.6}$ & ${1.39}_{-0.70}^{+0.74}$ & ${1.38}_{-0.91}^{+0.56}$ \\
082859.9+010756 & 14973 & ${0.51}_{-0.22}^{+0.62}$ & ${296}_{-98}^{+115}$ & ${2.5}_{-1.8}^{+1.7}$ & ${1.49}_{-0.80}^{+0.72}$ & ${0.56}_{-0.37}^{+0.44}$ \\
082952.7+002139 & 7528 & ${7.4}_{-6.9}^{+43.9}$ & ${59}_{-23}^{+41}$ & ${2.4}_{-1.7}^{+1.8}$ & ${1.62}_{-0.75}^{+0.62}$ & ${1.53}_{-1.02}^{+0.77}$ \\
082955.4+004131 & 3810 & ${8.0}_{-7.0}^{+42.6}$ & ${110}_{-42}^{+64}$ & ${2.5}_{-1.8}^{+1.8}$ & ${1.62}_{-0.76}^{+0.62}$ & ${1.37}_{-0.79}^{+0.56}$ \\
\end{longtable}
\tablefoot{Electronic version of the table is available at the CDS. Column 1: cluster name. Column 2: unique source ID presented in the eFEDS source catalog \citep{Brunner2021}. Columns 3, 4, 5, 6, 7: parameters of the \citet{Vikhlinin+06} model, $n_0^2$, $r_s$, $\epsilon$, $\beta$, $\alpha$ respectively.}

\end{document}